\documentclass[fleqn,usenatbib]{mnras}
\usepackage{newtxtext,newtxmath}
\usepackage[normalem]{ulem}

\usepackage[T1]{fontenc}

\DeclareRobustCommand{\VAN}[3]{#2}
\let\VANthebibliography\thebibliography
\def\thebibliography{\DeclareRobustCommand{\VAN}[3]{##3}\VANthebibliography}

\usepackage{graphicx}	
\usepackage{amsmath}	
\usepackage{todonotes}
\usepackage{microtype}
\usepackage{xcolor}
\usepackage{float}
\usepackage{comment}
\def\HI{\ion{H}{I}~}
\def\Hi{\ion{H}{I}}
\def\HII{\ion{H}{II}~}
\def\xHI{x_{\rm \ion{H}{I}}}
\def\xHII{x_{\rm \ion{H}{II}}}
\def\xb{\bar{x}_{\rm \ion{H}{I}}}
\def\xa{\bar{x}_{\rm \ion{H}{II}}}
\def\lya{Ly$\alpha$~}
\defcitealias{Lidz2007}{L07}	
\defcitealias{Xu2019}{X19}
\defcitealias{Worseck2014}{W14}	
\defcitealias{Bosman2022}{B22}

\usepackage{lipsum} 
\usepackage{comment}

\usepackage{tikz,xcolor,hyperref}
\definecolor{lime}{HTML}{A6CE39}
\DeclareRobustCommand{\orcidicon}{%
	\begin{tikzpicture}
	\draw[lime, fill=lime] (0,0) 
	circle [radius=0.16] 
	node[white] {{\fontfamily{qag}\selectfont \tiny ID}};
	\draw[white, fill=white] (-0.0625,0.095) 
	circle [radius=0.007];
	\end{tikzpicture}
	\hspace{-2mm}
}

\foreach \x in {A, ..., Z}{%
	\expandafter\xdef\csname orcid\x\endcsname{\noexpand\href{https://orcid.org/\csname orcidauthor\x\endcsname}{\noexpand\orcidicon}}
}

\title[The Forest at EndEoR]{The forest at EndEoR: The effect of Lyman Limit Systems on the End of Reionisation}

\author[I. Georgiev et al.]{
Ivelin Georgiev $^{1}$\thanks{E-mail: ivelin.georgiev@astro.su.se}\orcidA,
Garrelt Mellema$^{1}$\orcidB, and Sambit K. Giri$^{2}$\orcidC 
\\
$^{1}$Department of Astronomy and Oskar Klein Centre, AlbaNova, Stockholm University, SE-10691 Stockholm, Sweden\\
$^{2}$Nordita, KTH Royal Institute of Technology and Stockholm University, Hannes Alfvéns väg 12, SE-106 91 Stockholm, Sweden\\}

\date{Accepted in MNRAS. Received YYY; in original form ZZZ}

\pubyear{2024}

\begin{document}
\label{firstpage}
\pagerange{\pageref{firstpage}--\pageref{lastpage}}
\maketitle


\begin{abstract}
The final stages of cosmic reionisation (EndEoR) are expected to be strongly regulated by the residual neutral hydrogen in the already ionised regions of the Universe. Its presence limits the mean distance that ionising photons can travel and hence, the extent of the regions that sources of ionising photons can affect. The structures containing most of this residual neutral hydrogen are typically unresolved in large-scale simulations of reionisation. Here, we investigate and compare a range of approaches for including the effect of these small-scale absorbers, also known as Lyman limit systems (LLS), in such simulations. We evaluate the impact of these different approaches on the reionisation history, the evolution of the ultraviolet background, and its fluctuations. We also compare to observational results on the distribution of Lyman-$\alpha$ opacity towards the EndEoR and the measured mean free path of ionising photons. We further consider their effect on the 21-cm power spectrum. We find that although each of the different approaches can match some of the observed probes of the final stages of reionisation, only the use of a redshift-dependent and position-dependent LLS model is able to reproduce all of them. We therefore recommend that large-scale reionisation simulations, which aim to describe both the state of the ionised and neutral intergalactic medium use such an approach, although the other, simpler approaches are applicable depending on the science goal of the simulation.
\end{abstract}




\section{Introduction}

The Epoch of Reionisation (EoR) was the period in the first billion years of the Universe during which the neutral hydrogen (\HI) present in the intergalactic medium (IGM) was ionised by ultraviolet (UV) radiation. The ionising radiation is thought to originate from the first stars, galaxies, and accreting black holes, which formed due to the gravitational collapse of baryons into the dark matter halos. This phase transition spans a large range of physical scales. In addition to the cosmological properties of the large-scale IGM, the EoR is also defined by the astrophysical properties of galaxy formation and evolution \citep[see][for a review]{Pritchard_Loeb2012,dayal2018early}. 

A key constraint on reionisation comes from the measurement of the Thomson scattering optical depth ($\tau$) of Cosmic Microwave Background (CMB) photons \citep{Planck2020} against free electrons produced during the EoR. The scattering causes polarization of the CMB on scales larger than the horizon size and suppresses temperature anisotropies on scales smaller than the horizon size. The most recent constraint of $\tau =0.051 \pm 0.006$ suggests that the halfway point of the EoR was at $z_{\mathrm{mid}} \sim 7.5$ (assuming an instantaneous reionisation model).  Another key probe comes from the scattering of Lyman-$\alpha$ (Ly-$\alpha$, rest-frame wavelength of 1216 \AA), photons passing through an ionised region, by residual neutral hydrogen along a line-of-sight (LOS).  Notable observations include the \lya equivalent width \citep{Mason2018,Mason2019,Hayes2023, Hoag2019,Bolan2022,Bruton2023,Morishita2023,Jones2023}, the \lya dark pixel fraction \citep{McGreer2015,Jin2023} as well as the decreasing fraction of \lya emitters among Lyman break galaxies \citep{Ouchi2010,Endsley2022}. These can be utilised to place constraints on the global neutral fraction of hydrogen at a given redshift (see the top panel of Fig.~\ref{fig:1} in grey). 

Particularly the spectra of Quasi-Stellar Objects (QSO or quasars) at high-$z$ have been found to contain a wealth of information about the state of the IGM. Observations of the quasar damping wing \citep{Gunn-Peterson1965} in high-$z$ quasar spectra indicate the scattering of Ly-$\alpha$ photons by neutral hydrogen ($\Hi$) and can be used to estimate its fraction as a function of redshift \citep[e.g.][]{Mortlock2011,Greig2017,Banados2018,Davies2018b,Greig2019,Wang2020,Yang2020,Greig2022}. Therefore, QSOs are an excellent probe of the end stages of reionisation (hereafter EndEoR) \citep[see][for a comprehensive review]{Davies2023}. How the EndEoR occurred is essential for understanding the ionising budget of the universe \citep{Cain2023b}, and the role of the last islands of neutral hydrogen \citep{giri2024end}. While observations from \citet{Becker2001,Fan2006} initially indicated the endpoint of reionisation to be around $z \sim 6$, the \lya opacities such as those derived from QSO spectra in \citet{Bosman2022}
are difficult to reproduce with reionisation simulations, requiring an extended reionisation down to $z \approx 5.3$ \citep[see, e.g.][]{Kulkarni2019,Keating2020K}. 

Moreover, once a region of the IGM is re-ionised, its internal state depends on the ionised hydrogen (\HII) recombination rate against the ionisation rate set by the photon density or Ultraviolet Background (UVB) in the ionised IGM. How the mean UVB within such ionised areas changed over time depends on the growth and merging of such ionised structures. We will refer to the mean UVB value in the later stages of the EoR and post-EoR as the UV Background. Measurements of this quantity towards the EndEoR exist \citep{Calverley2011,Wyithe2011,Gaikwad2023} but most simulations struggle to reproduce this number, typically reporting UVB values higher than the measurements \citep[e.g.][]{Bolton2007,Finlator2009, Aubert2015,Dixon2016,Shukla2016,Ocvirk2020}. The value of the UVB depends on the mean free path (MFP) of ionising (Lyman continuum) photons, which is the mean distance an ionising photon travels before being absorbed \citep[e.g.][]{Fan2024}. The MFP can be set by the absorption of ionising photons by neutral regions at the edge of the ionised bubble or by absorptions within the ionised regions. Absorption within the \HII bubble could occur due to residual neutral hydrogen or due to the presence of discrete dense systems with an optical depth to ionising photons of $\tau > 1$, referred to as Lyman Limit Systems \citep[LLS, see e.g.][]{Alvarez2012, McQuinn2011, Shukla2016, D'Aloisio2020, Theuns2024,Fan2024}. The latter case is dominant in the post-reionisation Universe as fully neutral islands have disappeared. The MFP value towards the EndEoR has been measured by \citet{Becker2021,Zhu2023,Gaikwad2023} and found to be significantly smaller compared to the inferred value from post-EoR observations in \citet{Worseck2014} (hereafter W14). Therefore, the evolution of the MFP of ionising photons is particularly sensitive to the final stages of reionisation and vice versa, while its value is influenced by the presence of LLS.

Modelling the EndEoR has not received much attention in the literature, partly because the required simulation techniques are non-trivial but also because the prospective probe of the 21-cm signal from the IGM will be weaker. In reionisation simulations, the opacity of the IGM for ionising photons becomes essential, and there exists quite a bit of uncertainty about this quantity. For example, the excursion set approach used in {\sc 21CMFAST} \citep{Mesinger2007} is not applicable once most of the IGM has reionised. This limitation of the approach had triggered the development of {\sc ISLANDFAST} \citep[]{Xu2016}. This code begins reionisation of the IGM using the same method as in {\sc 21CMFAST} and later switches to another formalism developed in \citet{xu2014analytical} to follow the evolution of neutral hydrogen islands. 

In this work, we use the state-of-the-art {\sc} C$^2$Ray simulation framework that does not have this limitation as it numerically solves the three-dimensional radiative transfer equation. The main objective of this work is centred on studying the final stages of reionisation. Notably, we investigate the role of unresolved LLS, which we will also refer to as small-scale absorbers, in large-scale reionisation simulations. We construct and compare various common approaches to modelling the effect of these small-scale absorbers to examine their validity across a wide range of regimes. To achieve this, we test the implementation performance against multiple observables of reionisation, from probing the state of the ionised IGM in quasars to studying the neutral islands with the 21-cm signal from the IGM during the EndEoR. 

The paper is organised as follows. Sec.~\ref{sec:Theory} describes the simulations used within this study, focusing on the LLS implementations necessary for our analysis. In Sec.~\ref{sec:UVB}, we examine the role of the small-scale absorbers on the global parameters of reionisation and their effect on the statistics of the UVB background. In Sec.~\ref{sec:EoR_Observables}, we examine the observables at the EndEoR, such as the effective \lya optical depth, the mean free path of ionising photons, as well as the effect on the neutral islands at the end of reionisation through the lens of the 21 cm power spectrum. We summarise our findings and elaborate on future improvements in Sec.~\ref{Sec:Conclusions}. Unless stated otherwise, we denote comoving megaparsec as Mpc.

\section{Modelling}
\label{sec:Theory} 
In this section, we introduce our simulation framework and present the methods used to model the LLS.
\subsection{Reionisation Simulations}
\label{Sec:Simulations}

The methodology for producing our reionisation simulations was described in detail in \citet{iliev2006, Mellema2006b} and \citet{Dixon2016}. Here we only provide an outline. First, the density and halo fields are computed with the cosmological $N$-body code {\sc CUBEP$^{3}$M} \citep{Harnois2013}. Halos were found on-the-fly using the spherical overdensity halo-finder described in \citet{Watson2013}. Afterwards, these outputs are post-processed using the {\sc $\mathrm{C}^2$Ray} radiative transfer code \citep{Mellema2006a}.

The sources of ionising photons are associated with the halos. For the simulations used here, the ionising source model makes sure that the total number of ionising photons at a given redshift is always proportional to the fraction of matter which has collapsed into dark matter halos above the mass threshold, $f_{\mathrm{coll}}$. To achieve this, the rate of ionising photons escaping a halo of a given mass is calculated by multiplying the growth rate of halos by a constant $\zeta$, an emissivity parameter. The resulting value is then distributed over all halos above the minimal mass contributing ionising photons by assuming a linear relationship between the ionising luminosity and the mass of the halo \citep[see][for more details]{giri2021measuring}. This source model is described by the equation
\begin{equation}
\label{eq:emissivity}
    \Dot{N}_{\gamma} = \zeta \frac{M}{m_{\mathrm{p}}}\frac{ \Omega_{\mathrm{b}}}{ \Omega_{\mathrm{m}}} \frac{1}{f_{\mathrm{coll}}} \frac{\mathrm{d}f_{\mathrm{coll}}}{\mathrm{d}t} \ ,
\end{equation}
where $\Dot{N}_{\gamma}$ is the emissivity of a single source in units of s$^{-1}$, $M$ is the mass of the dark matter halo, $m_{\rm p}$ is the mass of a proton, and $\Omega_{\mathrm{m}}$ and $\Omega_{\mathrm{b}}$ are the density parameters for dark and baryonic matter. The simulations presented in Tab.~\ref{tab:C2sims} have a volume of $(L_{\mathrm{box}}\approx 349~\mathrm{cMpc})^3$ with $4000^{3}$ dark matter particles, resolving halos of mass down to $10^9 M_{\odot}$. The density field is gridded to a Cartesian 3D mesh with a size of $(N_{\mathrm{grid}}= 250)^{3}$ by applying a smooth-particle-hydrodynamics kernel \citep{Monaghan1992,Iliev2014}. We note that unlike the simulations described in \citet{Dixon2016}, no subgrid haloes and radiative feedback effects are used. We  have adopted a flat $\Lambda$CDM cosmology with parameters
($\Omega_{\mathrm{m}},\Omega_{\mathrm{b}},h,n_{s},\sigma_{8}$) = (0.27, 0.044, 0.7, 0.96, 0.8) in agreement with \textit{WMAP} \citep{hinshaw13} and \textit{Planck} results \citep{planck14}. 

We can express the relationship between the ionisation rate and the global emissivity ($\Dot{N}_{\mathrm{glob}}$, given in units of s$^{-1}$ cm$^{-3}$) \citep[see eq.10 of][]{Becker2021} as\footnote{The ``local source approximation'' is most valid for cases, where photons are immediately absorbed after emission and is less accurate during the EndEoR, when the MFP is large. Nevertheless, we apply this approximation as a rough estimate of the emissivity parameter needed to reionise the universe by redshift 6.}
\begin{equation}
    \label{eq::UVB_Becker21}
    \Gamma = \Dot{N}_{\mathrm{glob}} \epsilon \sigma_{\HI} \lambda_{\mathrm{MFP}},
\end{equation}
where $\sigma_{\HI} = 6.3 \times 10^{-18} \mathrm{cm}^{-2}$ is the hydrogen ionisation cross-section, $\epsilon = \alpha_{s} /(\alpha_{\mathrm{bg}}+2.75) \approx 0.5$ is the efficiency parameter and $\lambda_{\mathrm{MFP}}$ is the mean free path of ionising photons. Given the constraints of the UVB value by \citet{Calverley2011} and the MFP measurements of \citet{Becker2021,Zhu2023, Roth2023,Satyavolu2024}, we estimate an emissivity parameter of roughly $\zeta = 50$ is required for the simulations used within this body of work to have completed reionisation around $z = 6$. We use this value for all simulations in this paper.

\subsection{Modelling the effect of small-scale absorbers.} 
\label{subsec:LLS_models}
All simulations in this work, listed in Tab.~\ref{tab:C2sims}, are identical in terms of density fields, halo lists and source model. However, they differ in the implementation of the effect of small-scale absorbers to investigate the performance of each approximation during reionisation. In this section, we describe the five different ways in which we include their effect, comparing with and commenting upon the different approaches applied in the literature (see Tab.~\ref{tab:LLS_sims} for a non-exhaustive list).

The \texttt{r40} simulation models the MFP of ionising photos due to the presence of small-scale absorbers by limiting the distance photons are allowed to travel from a source. Such a spherical cut-off barrier is easy to implement, reduces the cost of the radiative transfer by limiting it to a certain distance and is also typically how semi-numerical codes based on the excursion set model of \citet{Furlanetto2004} implement the effect. Examples of such codes are the original {\sc 21CMFAST} code \citep{Mesinger2007}, {\sc simfast21} \citep{santos2010fast} and {\sc Sem-Num} \citep{majumdar2014use}.
Once a ray reaches the distance $R_{\mathrm{max}}$, its associated photons are assumed to have been absorbed by an unresolved LLS and are removed. For the \texttt{r40} simulation, $R_{\mathrm{max}} = 40$ cMpc was chosen across all redshifts. This approach, hence, is aimed at capturing the effect the LLS have on the global properties of reionisation on physical scales larger than $R_{\mathrm{max}}$, and naturally cannot capture the role of small scale absorbers below the cut-off scale. Improvements to the method, such as those in \citet{giri2024end} introduce an evolving spherical barrier model, to account for the redshift evolution of LLS.

The \texttt{C2} simulation instead models the effect of small-scale absorbers by boosting the recombination rate through the introduction of a clumping factor, $C = \langle \rho^{2}\rangle/ \langle \rho \rangle^{2}$ where $\rho$ is the density of matter. The choice of global clumping factor is motivated by the common conception that it is sufficient to model small-scale absorbers \citep[][and references within]{Wu2021}, as most ionising photons are absorbed in the fully ionised (and hence recombining) outer layers of self-shielded systems. Crucially, introducing a clumping factor in the radiative transfer equations allows for a self-consistent relationship between the ionisation and recombination rates, directly affecting properties such as the UVB and the MFP of ionising photons. Here we use a simple global clumping factor and set its value to $C=2$ at all times, though there is an indication of its dependence on redshift and the global mass-weighted ionisation fraction \citep[see][]{Chen2020}. Note that this clumping comes in addition to the clumping already resolved within the simulation from the baryon density fluctuations. In order to limit the computational cost and stop photons from crossing more than half a volume size, the \texttt{C2} simulation also includes a cut-off barrier of R$_{\mathrm{max}} = 70$ cMpc. The choice of the cut-off is motivated by the MFP of ionising photos at the post-EoR reported by \citetalias{Worseck2014} at $z = 5$. 

The clumping factor is the approach most widely explored in the literature. \citet{Mao2020} compared the use of global clumping factors and density-dependent ones in simulations similar to the ones we use here and showed that global clumping factors are a reasonable approximation for density-dependent clumping. \citet{Bianco2021} showed that the impact of stochasticity on the density-dependent clumping is relatively small. Notably, \citet{Cain2023a} compared a sub-grid clumping model calibrated to radiative hydrodynamics simulations to methods commonly used in the literature, such as a global clumping factor
and a dynamic model which includes pressure smoothing of the IGM. Their findings suggest slight variations in the bubble topology during reionisation, yet stark differences between the models are most pronounced when varying the source models, particularly when the EoR is driven by bright sources. Such studies demonstrate that while prescriptive clumping models can adequately be applied in studies of the neutral IGM, modelling its ionised counterpart calls for a more nuanced approach.

\begin{table}
\centering
\caption{{\sc C$^2$-Ray} simulations with a minimal mass of ionising sources of $M_{\mathrm{min}} = 10^9 M_{\odot}$, an emissivity parameter of $\zeta =50$, and their respective models of the LLS.} 
\label{tab:C2sims}
\begin{tabular}{c|c|c} 

Label & MFP Method & $\lambda_{\mathrm{MFP}} $ \\ \hline
\texttt{r40} & spherical cut-off barrier& 40 cMpc
\\ \hline
\texttt{C2}  & global clumping of  C = 2 & 70 cMpc
\\ \hline
\texttt{eLLS}  & uniform absorption & equation~\eqref{eq:W14fit}
\\ \hline
\texttt{pLLS}  &position-dependent absorption & equation~\eqref{eq:W14fit}
\\ \hline
\texttt{LLS40} & uniform absorption& 40 cMpc
\end{tabular}
\end{table}

\begin{table*}
\centering
\caption{Table of approaches implement when modelling small-scale absorbers across the literature.} 
\label{tab:LLS_sims}
\begin{tabular}{c|c} 
spherical cut-off barrier & \parbox{0.7\textwidth}{{\sc 21CMFAST} \citep{Mesinger2007}, {\sc simfast21} \citep{santos2010fast}, {\sc Sem-Num} \citep{majumdar2014use}, {\sc py$\mathrm{C}^2$Ray} \citep[evolving barrier model][]{Hirling2024}.} \\ \hline
clumping factor models & \parbox{0.7\textwidth}{{\sc $\mathrm{C}^2$Ray} \citep{Mao2020,Bianco2021}, {\sc SCORCH} \citep{Chen2020}, {\sc CROC} \citep{Kaurov2014}, {\sc LORELI} \citep{Meriot2024}, \citep{Cain2023a}, {\sc FlexRT} \citep{Cain2023a}., {\sc SCRIPT} \citep{Maity2022}} \\ \hline
MFP/additional opacity  & \parbox{0.7\textwidth}{ {\sc $\mathrm{C}^2$Ray} \citep{Shukla2016, Georgiev2022}, {\sc 21CMFAST} \citep{Davies2021b}, {\sc FlexRT} \citep{Cain2021}, {\sc AMBER} \citep{Trac2022}, {\sc IslandFAST} \citep{Xu2017}, {\sc Astreus} \citep{Hutter2018}. } \\ \hline
inhomogeneous recombinations  & \parbox{0.7\textwidth}{ {\sc 21CMFAST} \citep{Sobacchi2014}, {\sc IslandFAST}\citep{Zhu_Xu_2023}} \\ \hline
\end{tabular}
\end{table*}
The three remaining simulations model the effect of small-scale absorbers by adding additional, subgrid, opacity for ionising photons, characterized by a parameter $\lambda_{\mathrm{MFP}}$, the distance at which this additional opacity corresponds to an optical depth $\tau_\mathrm{LL}$ of 1 for Lyman limit photons. This additional opacity is frequency-dependent following the frequency-dependence of the cross-section of neutral hydrogen, approximately $\nu^{-3}$. A value of $\tau_\mathrm{LL}=1$ implies that around 63 per cent ($1-e^{-\tau}$) of the photons at the Lyman limit will have been removed by the subgrid opacity at $\lambda_{\mathrm{MFP}}$. For higher frequency photons this fraction will be lower. Note that this is an additional opacity as the neutral hydrogen on the grid will also remove ionising photons. Unlike the cut-off barrier $R_\mathrm{max}$ model, absorption due to small-scale structure is not instantaneous at a fixed distance from the source, but occurs gradually at all distances from a source. We therefore will sometimes refer to it as a ''diffuse barrier'' model, although no real barrier exists in this approach. The small-scale absorbers are distributed throughout the IGM according to the method previously described in \citet{Shukla2016}
and in the appendix of \citet{Georgiev2022}. This approach is similar in philosophy to the one described in \citet{Davies2021}, who directly modify the excursion formalism by including the additional attenuation below R$_{\mathrm{{max}}}$, which is modulated by the MFP of ionising photons inferred from fits \citep[see][for an example]{Songaila2010}.
Note that the approach applied here differs from the methods described in \citet{Xu2017}, \citet{Hutter2018}, \citet{Trac2022}, where a fit of $\lambda_{\mathrm{MFP}}$ is directly applied when calculating the ionisation field. Moreover, the diffuse models do not account for the dynamic response of IGM clumping and LLSs to the EoR \citep[see][for a subgrid model]{Cain2021}.

Simulations \texttt{LLS40}, \texttt{eLLS} and \texttt{pLLS} all use this ''diffuse barrier'' model for the effect from small-scale absorbers. The difference is that in \texttt{LLS40}, $\lambda_{\mathrm{MFP}}$ is uniform and fixed at 40 cMpc for all redshifts. In \texttt{eLLS}, $\lambda_{\mathrm{MFP}}$ is also uniform but evolving according to
\begin{equation}
    \lambda^\mathrm{W}_{\mathrm{MFP}}=37 \mathrm{\,pMpc}\left(\frac{1+z}{5}\right)^{-5.4}\,,
    \label{eq:W14fit}
\end{equation}
a fit taken from \citet{Worseck2014}. As $\lambda^\mathrm {W}_{\mathrm{MFP}}\approx$ 40~cMpc at z=6, the \texttt{LLS40} and \texttt{eLLS} simulations have the same MFP at that redshift. The \texttt{pLLS} simulation uses the same evolving $\lambda^\mathrm {W}_{\mathrm{MFP}}$ fit to determine the average subgrid opacity, but distributes the opacities proportional to the sum of the virial cross-sections of all resolved halos (including those hosting ionising sources) in a cell, which makes the subgrid opacity position dependent. It assumes that the small-scale absorbers are mostly associated with dark matter halos \citep[see][]{Theuns2024}. This approach was described in more detail in sec.~3.3 of \citet{Shukla2016}. 

Table~\ref{tab:LLS_sims} also lists a fourth type of approach, denoted as ''inhomogeneous recombinations''. This approach also uses a clumping factor but includes the effects of self-shielding on the local subgrid density fluctuations and the level of the local ionization rate to determine the effective local clumping factor. In this paper we do not include such a local and dynamic approach but plan to do so in future work.
\begin{figure}
\includegraphics[width=\columnwidth]{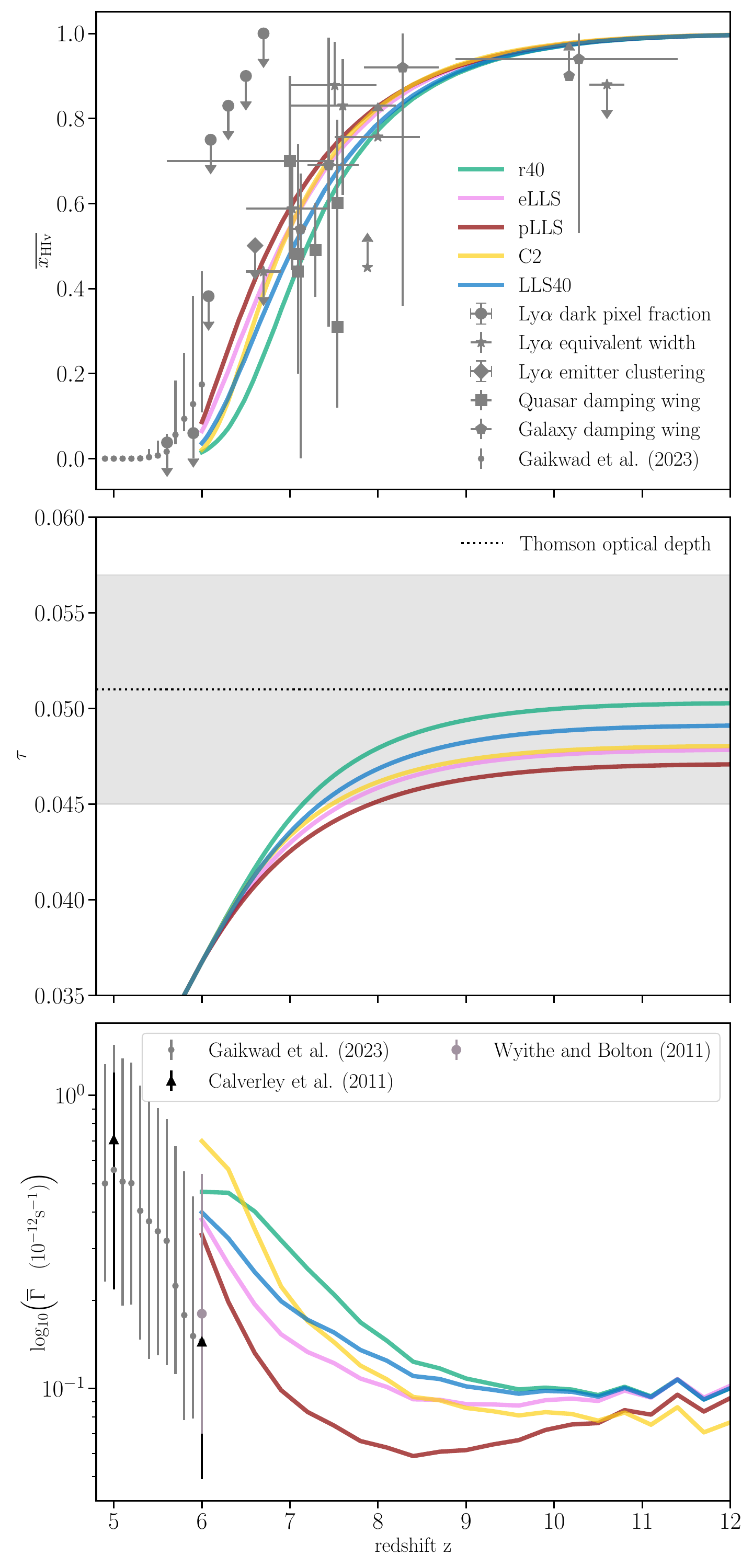}
\caption{Top Panel: Reionisation histories for the models in Table \ref{tab:C2sims}. Constraints on the ionisation histories from \lya equivalent width (stars) \citep{Mason2018,Mason2019,Hoag2019,Bolan2022,Bruton2023,Morishita2023,Jones2023}, \lya dark pixel fraction (circles) \citep{McGreer2015,Jin2023}, and \lya emitters (diamonds) \citep{Ouchi2010} as well as the damping wings from quasars (squares) \citep{Greig2019,Yang2020,Greig2022} and galaxies (pentagons)\citep{Hsiao2023,Umeda2023}. Middle panel: The Thomson optical depth constraints as reported by \citet{Planck2020}. Lowest Panel: The evolution of the UVB background, compared against the measurements from \citet{Calverley2011},\citet{Wyithe2011}, and \citet{Gaikwad2023}. The inclusion of LLS sub-grid models both delays the final stages of the EoR and regulates the evolution of the UVB. Moreover, boosting the clumping factor throughout the simulation (\texttt{C2} model, in yellow) results in overshooting the UVB at EndEoR, whilst a low cut-off of the MFP (\texttt{r40} model, in green) limits its evolution.} 
\label{fig:1}
\end{figure}
\section{Effect of small-scale absorbers on reionisation and the ultraviolet background}
\label{sec:UVB}
In this section, we will compare the results of the five simulations in terms of their reionisation history and the evolution and statistical properties of the UVB.

\subsection{The global parameters of reionisation}
\label{subsec:global_params}
In Figure~\ref{fig:1}, we plot the global observables characterising our five reionisation models. This figure also introduces the colour scheme for these simulations which we will use throughout this paper. The top panel shows the evolution of the volume-averaged neutral fraction $\xb$ with observational constraints \citep[see][]{Bruton2023,Keating2023}. All simulations are consistent with these observational constraints. The plot also shows that of the five simulations only \texttt{r40} completes reionisation by redshift 6. Due to limitations of the $N$-body results, we were unable to continue the reionisation simulations beyond $z=6$. 

The middle panel shows the CMB Thomson optical depths against the measurement of $\tau= 0.051 \pm 0.006$ from \citet{Planck2020}. The reionisation histories of all models are consistent with those measurements. 

In the bottom panel of Fig.~\ref{fig:1}, we plot the average UVB as it evolves with redshift during the EoR. We calculate the average UVB by normalising the ionisation rate output from our simulations with respect to the ionisation fraction ($\xHII$) in each cell following $\sum_{i} \Gamma_{i} x_{\mathrm{H_{II}},i} / \sum_{i} x_{\mathrm{H_{II}},i}$. The measurements of the UVB shown are as reported by \citet{Calverley2011,Wyithe2011, Gaikwad2023}. This plot shows considerable differences in the evolution of the UVB, which we will now discuss in some detail.

How LLS are modelled in our simulations directly dictates the evolution of the average UVB. First, we consider the \texttt{C2} model (shown in yellow), which has a global clumping factor of $C=2$ and cut-off scale of $R_{\mathrm{max}} = 70$ cMpc. The addition of a global clumping parameter results in a boost of the recombination rate \citep[see][]{Bianco2021}. A large value of the clumping factor will, therefore, impact the UVB and be more pronounced in denser regions of the simulation. This becomes evident by examining the evolution of the UVB for the \texttt{C2} model compared to the remaining models. The first ionising sources are located in the regions of the highest density. The clumping factor will increase the already large recombination rates there and thus limit the growth of these initial $\HII$ regions. This initially delays the EoR, evident in the neutral faction's slower evolution and the lower UVB amplitude at $z >8$. Despite this, below $z = 8$, the clumping model shows the strongest evolution in the UVB and overshoots the observations at $z = 6$. This is due to two key factors. After $\xb \approx 0.8$, the ionised bubbles expand past their local over-densities and into the lower density IGM, in addition to a considerable overlap of ionised regions \citep{Lidz2008}. Regions previously in equilibrium with the ionisation output of their local sources are affected by the radiation of adjacent sources, boosting the UVB. Moreover, as $R_{\mathrm{max}} = 70$ cMpc for the \texttt{C2} model, the ionising radiation can in principle propagate further than in our other models. This behaviour becomes significant past the overlap phase at the end stages of reionisation when most cells within the simulation host ionising sources and the remaining neutral islands reside in the deepest voids. 

The opposite effect on the UVB can be seen when considering the \texttt{r40} model, which has no additional clumping but a cut-off spherical barrier of $R_{\mathrm{max}} = 40$ cMpc. Compared to \texttt{C2}, the cut-off barrier model (seen in green in the bottom panel of Fig.~\ref{fig:1}) initially has a higher UVB 
value, which increases during reionisation and eventually asymptotes around $z \approx 6.3$. During the early phases reionisation in \texttt{r40} is more rapid than in \texttt{C2} as there is no additional clumping to impede the growth of the \HII bubbles. However, towards the end stages of the EoR, the characteristic scale of \HII bubbles approaches the scale of the cut-off barrier, which impedes the further growth of these regions\footnote{Note that \HII bubbles will continue to grow in size due to overlap, yet no single source can create an ionised structure larger than the cut-off barrier \citep[see appendix of][]{Georgiev2022}.}. Moreover, towards the end of reionisation, most neutral islands are far away from ionising sources, as they are located in the voids \citep{Giri2019, giri2024end}. The sharp cut-off of  40 cMpc results in fewer ionising photons reaching the outskirts of the neutral islands. Therefore, once a neutral island becomes ionised, its local ionisation rate is lower, and the residual neutral fraction is higher than in models with larger $R_{\mathrm{max}}$ values. Because of this, most recently ionised cells in these models will have a lower ionisation rate, pushing down the global mean.
In addition to \texttt{r40}, we have run models with lower cut-off barriers of 10 and 20 cMpc to confirm this effect. While low values of the cut-off barrier can be motivated by observations, the MFP is the mean value around a broad distribution, and in this approach, the free paths larger than the cut-off barrier are not modelled. We argue that this will also impact the evolution of the UVB. 

For all the diffuse barrier models, the UVB is slower to evolve than in \texttt{r40}, and reionisation is delayed \citep[also see fig.2 of][]{Shukla2016}. The redshift-dependent opacity from the LLS in \texttt{eLLS} (in fuchsia) further slows the reionisation process by impeding the growth of \HII bubbles compared to \texttt{LLS40}. The addition of the position and density dependence in \texttt{pLLS} (in red) further exacerbates this effect such that regions with more halos will have a higher opacity due to LLS in the voids of the simulation. However, as the EoR progresses and ionised bubbles overlap more and more, the evolution of the mean UVB for \texttt{LLS40}, \texttt{eLLS}, and \texttt{pLLS} exhibit a steeper gradient compared to \texttt{C2} and \texttt{r40} at the EndEoR and converge at $z = 6$. This behaviour follows the same line of thinking as the role of the global clumping parameter. However, because the additional opacity removes ionising photons, assumed to be absorbed by unresolved LLS, it does not affect the resolved recombination rate in our models.\footnote{In fact, the removed photons would indeed ionise the dense gas surrounding the halos, boosting the UVB in these regions and decreasing their absorption cross-section. We approximately account for this in our \texttt{eLLS} and \texttt{pLLS} models through the assumed evolution of the opacity due to LLS.} Therefore, it does not result in a boost of the UVB as seen in the \texttt{C2} model. This effect is most pronounced in the \texttt{pLLS} model since the additional position and density-dependent opacity causes dense regions to be more optically thick to ionising photons. On the other hand, low-density regions which host fewer halos will have less additional opacity added for the \texttt{pLLS} model, compared to  \texttt{LLS40} and \texttt{eLLS}. Because of this difference, the reionisation history seen at the top panel of Fig.~\ref{fig:1} is initially gradual for \texttt{pLLS} but becomes steeper at the end of reionisation.

In summary, for the \texttt{C2} model the clumping initially suppresses the evolution of the UVB, however, it overshoots its amplitude compared to the measurements at $z=6$. The role of the absorption due to clumping diminishes as the overlap of \HII regions becomes significant and UV radiation is free to propagate up to 70 cMpc. Meanwhile, the diffuse barrier models allow a fraction of the ionising photons to permeate past the defined MFP while the \texttt{r40} model imposes a firm upper limit. For the \texttt{r40} case, the sizes of \HII bubbles are sharply constrained by the imposed cut-off, and the UVB exhibits an asymptotic trend towards the EndEoR.

To conclude the discussion of the evolution of the mean UVB we would like to point out that the observed trends in the diffuse barrier models are more consistent with the UVB evolution reported in high-resolution state-of-the-art hydrodynamic simulations. This can be seen by comparing to the results shown in fig.~5 of \citet{Garaldi2022} ({\sc THESAN}) and fig.~1 of \citet{Lewis2022} ({\sc CODA III}). However, it should be remembered that those simulations consider volumes which are around $(4)^3$ times smaller than our simulations and hence do not sample the same range of environments.

\begin{figure*}
\includegraphics[width=.49\linewidth]{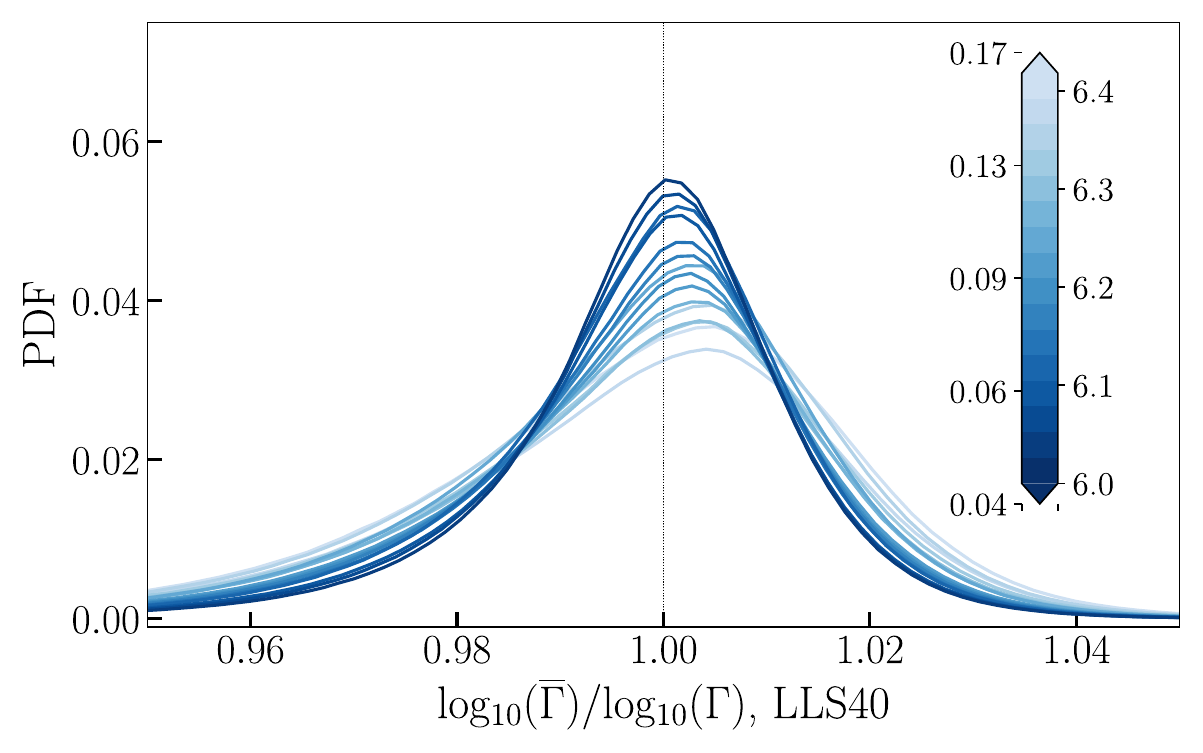}
\includegraphics[width=.49\linewidth]{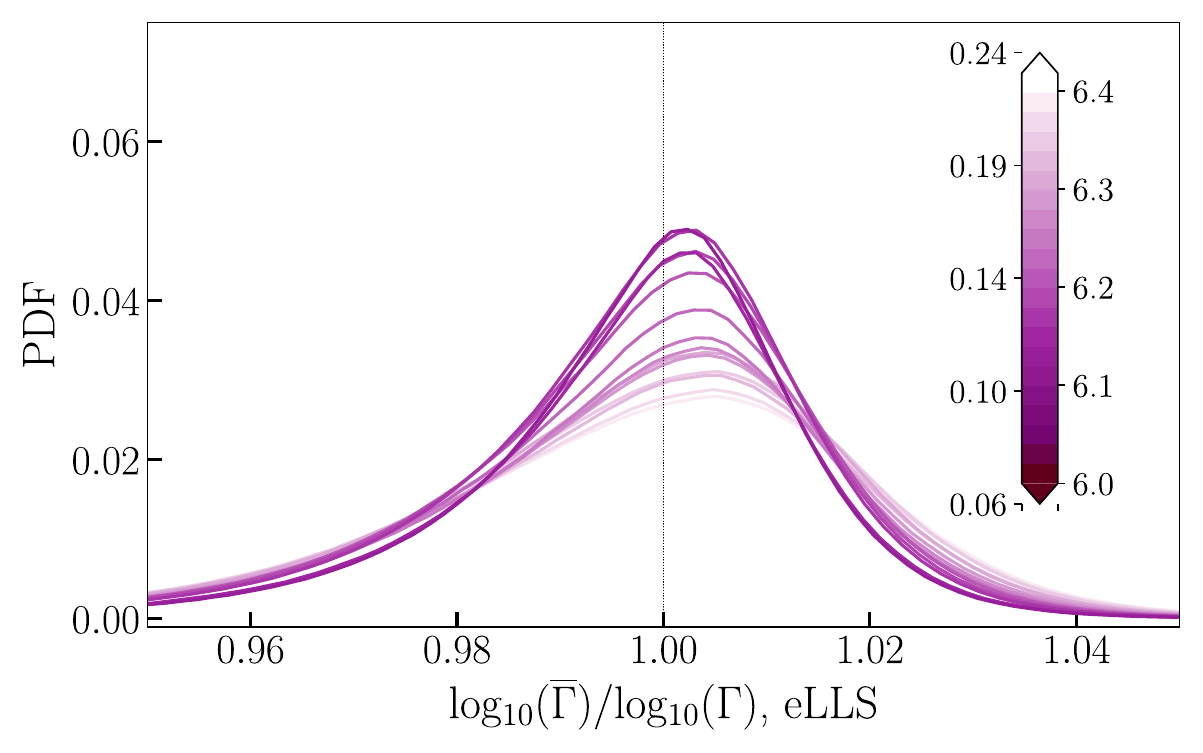}
\includegraphics[width=.49\linewidth]{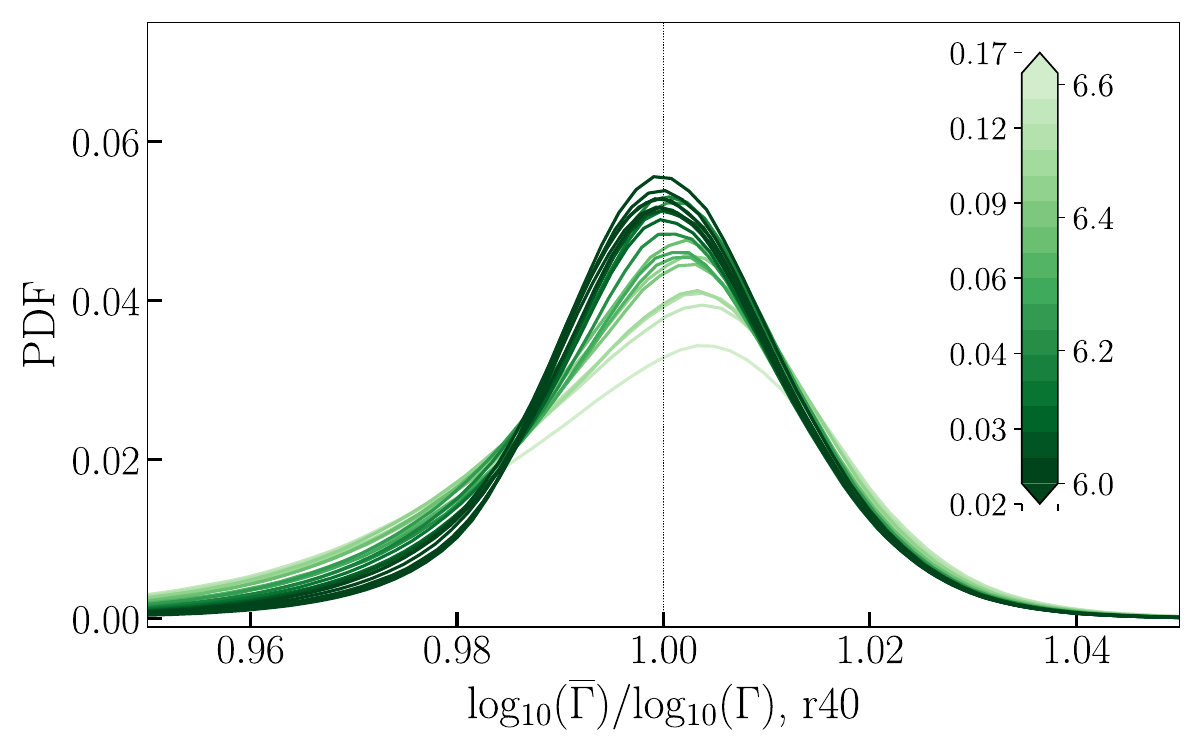}
\includegraphics[width=.49\linewidth]{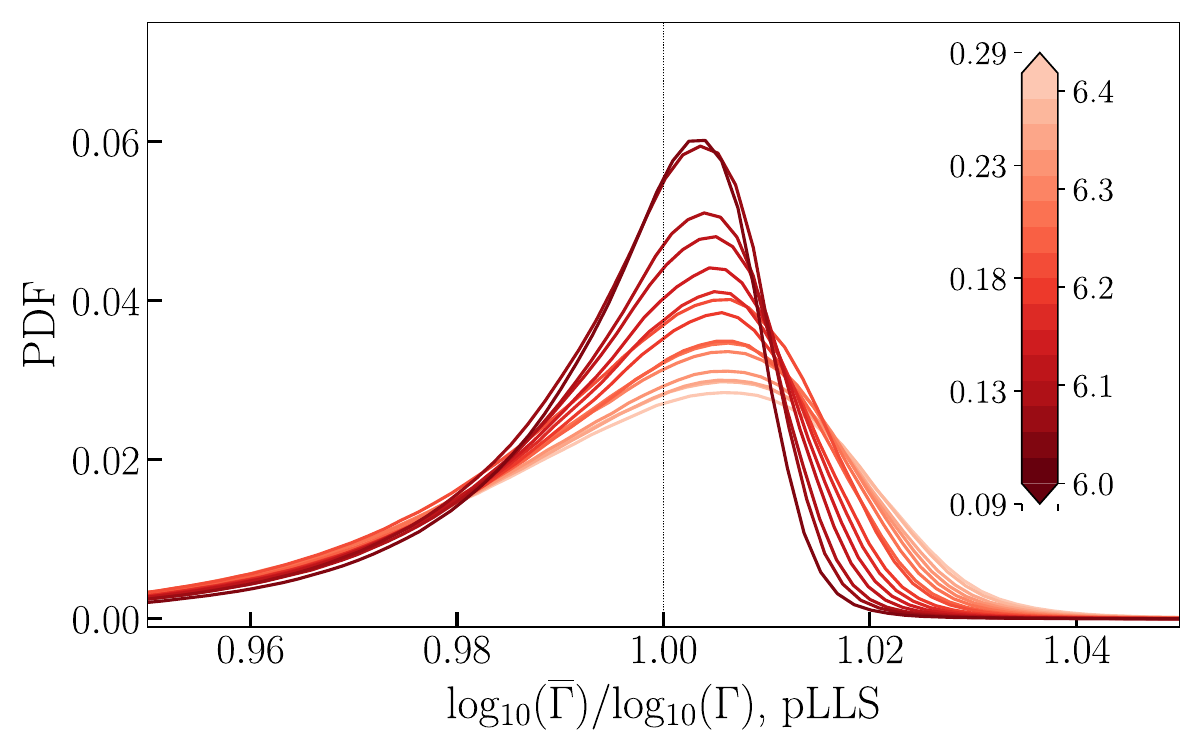}
\includegraphics[width=.49\linewidth]{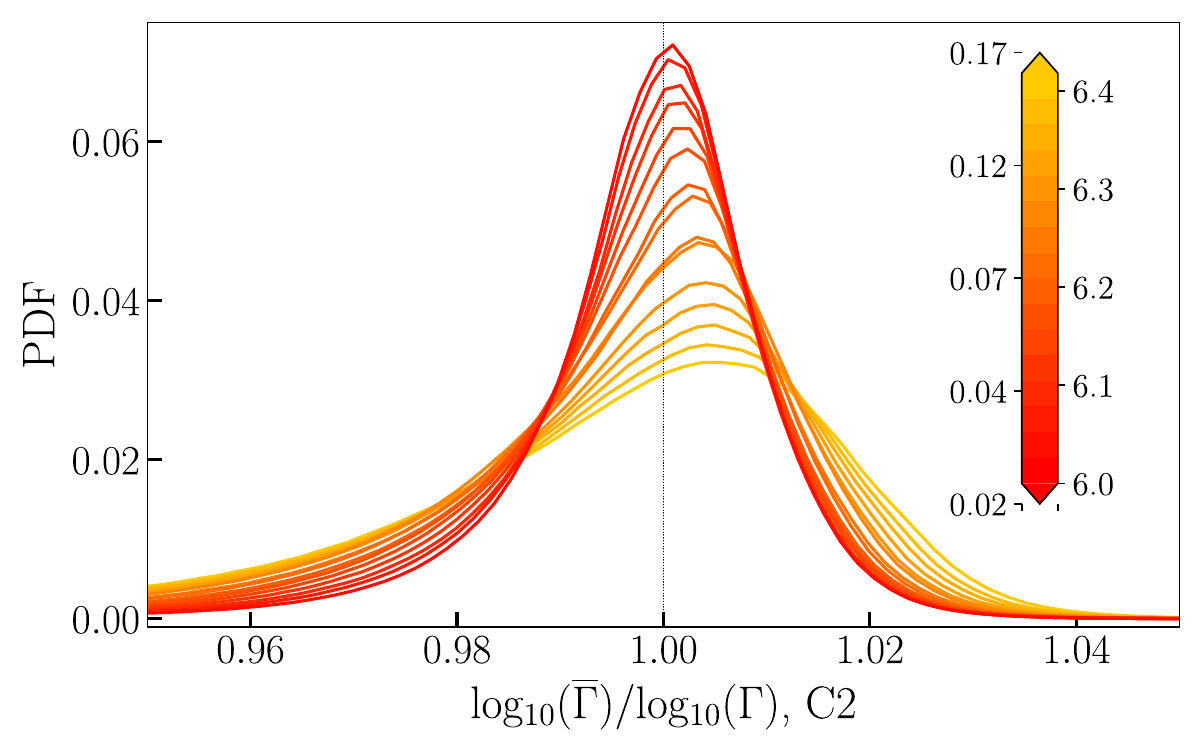}
\includegraphics[width=.49\linewidth]{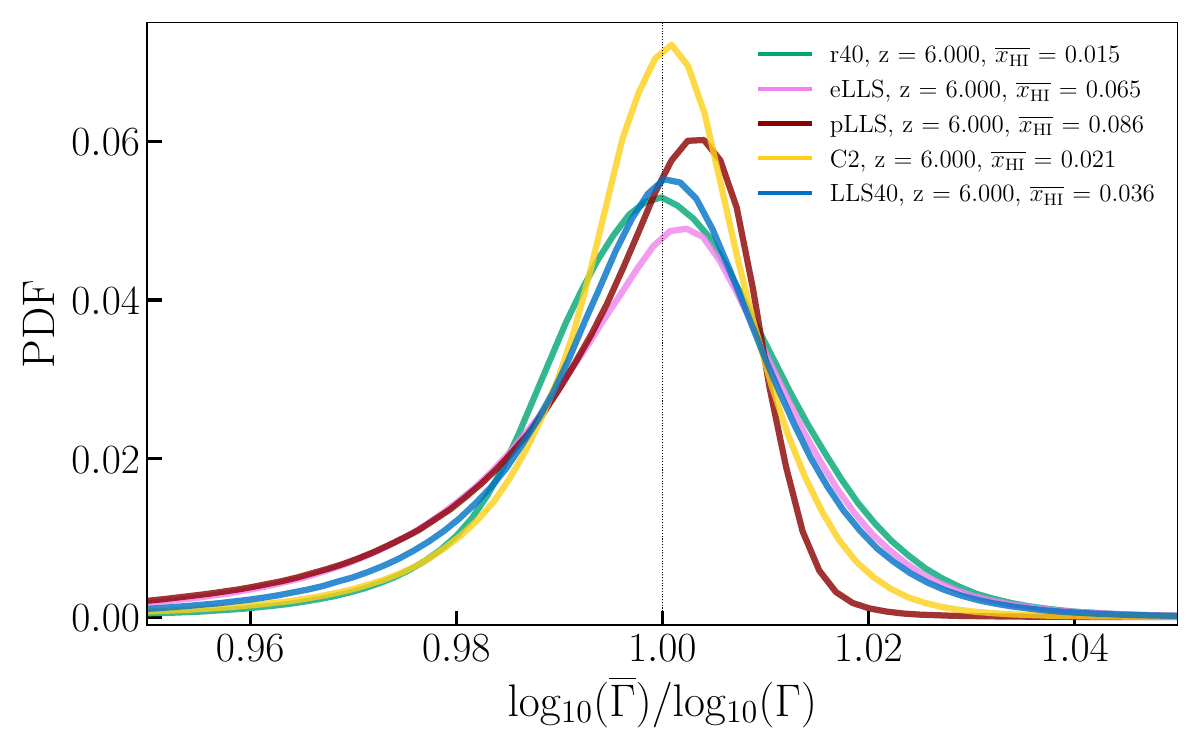}
\caption{PDFs of the logarithm of the UVB for ionised cells ($\xHII \geq 0.5$) for the set of models showcased in Tab.~\ref{tab:C2sims}. Note that UVB is normalised against the mean of the distribution at the given redshift. Values at the right of each colour bar are the redshift, whilst those to the left are the volume-averaged neutral fraction. The figure at the bottom right row includes the UVB statistics for each model at a fixed redshift of 6. A value of unity is plotted as a vertical line, to highlight the peak of each PDF. The role of the LLS model directly influences the shape of the UVB PDFs. The right tail of each PDF (values larger than the mean) is sensitive to small-scale fluctuation induced by the implementation of the small-scale absorbers. The effect is most pronounced in the \texttt{pLLS} model (middle row, right column, and seen in red), where the LLS are both position and redshift-dependent. Conversely, the left tail of each PDF (values smaller than the mean) is sensitive to the cumulative effect of the LLS models, i.e. the overall MFP. For example, the \texttt{r40} model (middle row, left column, and seen in green) imposes a fixed barrier of 40 cMpc, freezing the evolution of the large-scale UVB fluctuations (note each PDF is centred around the mean value at the given redshift). }
\label{fig::pdf}
\end{figure*}
\begin{figure*}
    \includegraphics[width=.99\linewidth]{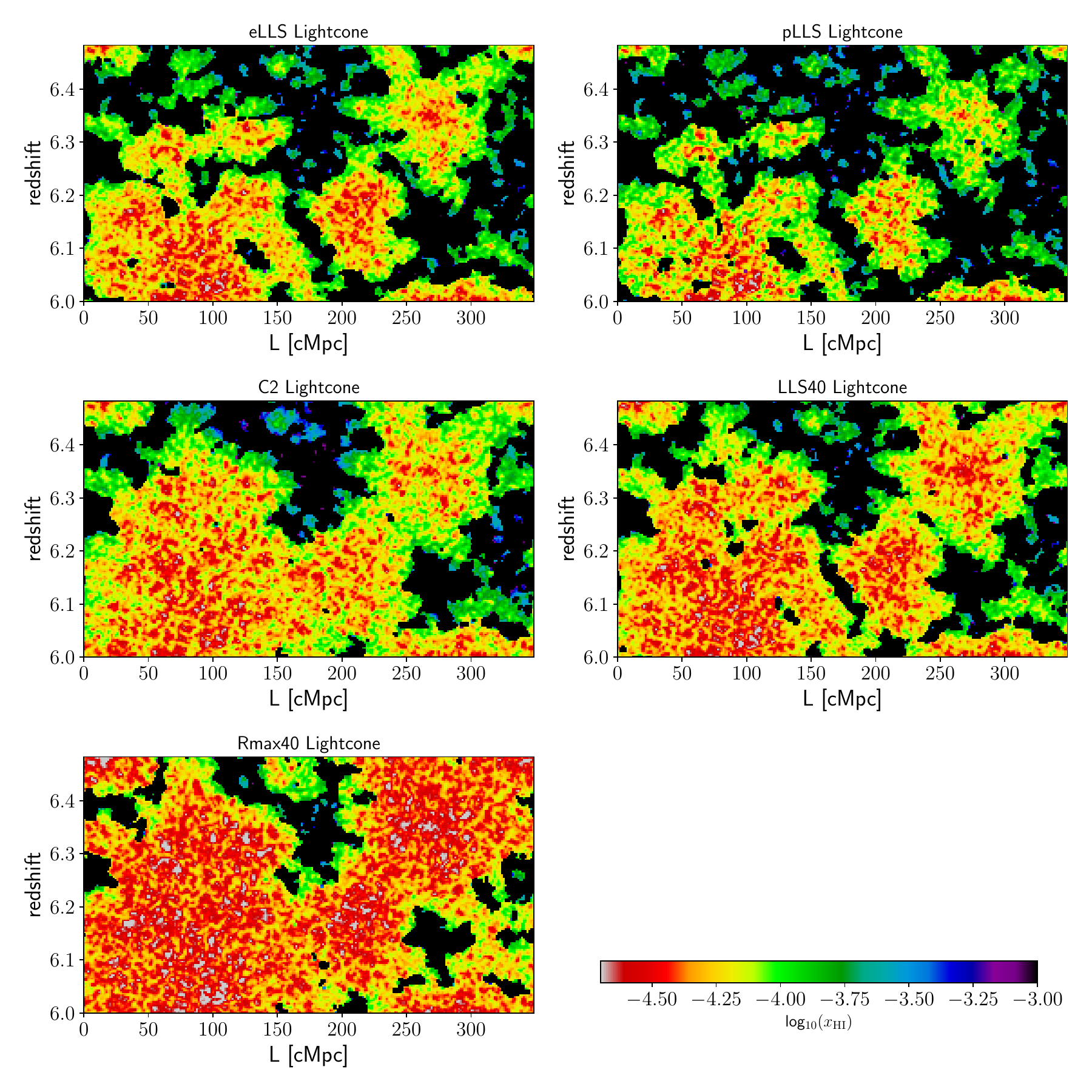}
    \caption{Light cones of the reionisation models presented in Tab.~\ref{tab:C2sims} during the EndEoR, where the $y$-axis for each panel covers a redshifts period lower than $z = 6.6$, while the $x$-axis is the length of simulation. The neutral fraction is plotted in a logarithmic scale, the colour scheme is used to highlight the fluctuations of the ionised and neutral IGM. For example, the dark black regions are indicative of neutral hydrogen islands, and the red regions indicate the ionised medium, particularly the residual neutral hydrogen. Fluctuations in the ionised IGM are pronounced in shades of yellow and green, highlighting the role of the small-scale absorbers. Because in ionisation equilibrium the ionisation rate scales as $\Gamma \sim 1/x_{\mathrm{HI}}$ these panels also provide a good indication of the UVB fluctuations shown in Fig.~\ref{fig::pdf}.} 
    \label{fig:LC}
\end{figure*}
\subsection{Statistical analysis of the UVB}
\label{sec::UVB_PDF_analysis}
In this section, we expand our analysis of the role of small-scale absorbers on the UVB from Sec.~\ref{subsec:global_params} by going beyond the single average value and instead examine the Probability Density Function (PDF) of the field. In Fig.~\ref{fig::pdf}, we present the PDF of the logarithm of the ionisation rate for all ionised cells ($\xHII \geq 0.5$). The PDF has been normalised against the logarithm of the mean UVB value at the given redshift to highlight the evolution of the fluctuations (i.e. $\log_{10}(\overline{\Gamma})/\log_{10}(\Gamma)$). Hence a ratio above unity is indicative of a UVB larger than the global mean and vice versa. As we saw above, while all the source models used in this paper are identical, the various implementations of the effect of small-scale absorbers result in a diverse range of reionisation histories. We present the PDFs for each model for the final stages of reionisation (roughly $z \leq 6.6$ and $\xa \geq 0.8$). Moreover, we include light cone slices of the neutral fraction of hydrogen for our models in Figure~\ref{fig:LC}. Compared to coeval slices of the simulation, where both axes are dimensions of length, the light cones
show the evolution of the neutral fraction across a redshift period over the length of the simulation, as the $y$ and $x$-axis, respectively. 
This serves to better visualise the role of small-scale absorbers over time, specifically in the case of the redshift-dependent models \texttt{eLLS} and \texttt{pLLS}. 

All models exhibit similar PDF distributions at high redshifts (and, therefore, high global neutral fractions), as seen in the lightest shades in Fig.~\ref{fig::pdf}. The distributions are initially skewed to values higher than the UVB mean, while there are significant outliers of the PDFs for values lower than the UVB mean. In this period, while reionisation is underway, the remaining neutral islands in the cosmic voids significantly affect the ionisation rates in low and high-density regions alike and, therefore, the fluctuations in the UVB. 

For example, we can imagine two adjacent regions, one of high and one of low density, separated by an area of neutral hydrogen. The high-density region within the simulation will likely host multiple clusters of ionising sources. Its local UVB value will be higher than that of the region with fewer sources (for example, see the lower left area of all panels in Fig.~\ref{fig:LC}). The lower-density region can be in a state of ionisation equilibrium if it is sufficiently shielded by the neutral islands. Naturally, as the neutral island in our example disappears with the progress of EoR, and the two regions become interconnected, radiation from the cluster of sources interacts with the IGM in the low-density region, pushing the local UVB to evolve to the global mean, reducing the fluctuations in the UVB distribution. In this regime, the role of small-scale absorbers (and how they are modelled) is most evident in the UVB fluctuations in the PDFs. The LLS model not only limits the global limit of how far ionising photons can travel (the MFP), but it also affects the UVB fluctuations inside the ionised region (i.e. small scales). 

We can observe multiple trends due to the nature of each method of modelling the small-scale absorbers. We first discuss the sides of the PDFs of our models throughout the EndEoR. The shift of the peak of the PDFs with a redshift to their mean value is indicative of the evolution of the ionisation rate to a state of equilibrium. Moreover, we can initially associate the right side of the PDF with high-density regions inside ionised bubbles, occupied by numerous sources, which in the case of the \texttt{pLLS} model are sensitive to the presence of small-scale absorbers. Meanwhile, the left side of the PDF is more sensitive to the low-density regions, which interact with the ionisation font, which is modulated by the MFP.

For the models \texttt{r40,LLS40,C2} in the left column of Fig.~\ref{fig::pdf}, we note that the outliers on the left side of each PDF decrease with redshift, indicating the reduction of large-scale fluctuations in the UVB. The common factor between the diffuse barrier (blue model), the cut-off barrier (green model), and the clumping model (orange model) is that they do not directly model the small-scale absorbers, resulting in a more homogeneous local UVB field, which adjusts to the global UVB value as overlap occurs. This is especially true for the \texttt{r40} and \texttt{LLS40}, which model the LLS by placing a non-evolving upper limit on the MFP. On the other hand, the clumping factor in the \texttt{C2} model accounts for the position-dependence of small-scale absorbers. Rather than being absorbed by the self-shielded regions photons are free to propagate across the simulation, leading to an increased UVB value as discussed in Sec.~\ref{subsec:global_params}.
Conversely, the models on the right column of Fig.~\ref{fig::pdf} (\texttt{eLLS,pLLS}), exhibit large-scale fluctuations, which do not decrease with the EndEoR. For the redshift and density-dependent model \texttt{pLLS}, this is a natural consequence of adding the small-scale absorbers, which modulate the large-scale UVB. Adding a redshift dependence of the LLS evolution in the \texttt{eLLS} model induces this effect compared to the \texttt{LLS40} model, as its increasing MFP value limits the large-scale UVB fluctuations throughout reionisation. 

We now turn our attention to the fluctuations seen on the right side from the mean value of each PDF\footnote{Remember that the PDFs are centred around their evolving mean value.}, which sre sensitive to the small-scale fluctuations of the UVB and the presence of the small-scale absorbers. The previous point is now more apparent by comparing the \texttt{r40} model against the \texttt{LLS40, eLLS}, and \texttt{pLLS}. The cut-off barrier model places an upper bound on the MFP. Hence, all small-scale fluctuations are smoothed over within a given region. Thus, the PDF's right-hand side does not significantly evolve compared to the mean value. A diffuse barrier model, such as in \texttt{LLS40}, removes more local photons due to the absorption by self-shielded regions. Additionally, the diffuse barrier allows photons from distant sources to propagate further than those in the cut-off barrier in \texttt{r40}. Moreover, the evolving diffuse barrier in \texttt{eLLS} reproduces this effect and induces large-scale fluctuations in the UVB.

The most interesting effect can be observed in the \texttt{pLLS} model, as compared to the clumping model \texttt{C2}. By accounting for the density/position and redshift dependence of the LLS, these systems induce small-scale UVB fluctuations and regulate the large-scale UVB (as seen in the top right panel of Fig.~\ref{fig:LC}, considering that in equilibrium $\Gamma \sim 1/x_{\HI}$). The resulting PDF closely follows a log-normal distribution. While the clumping model induces small-scale fluctuations, as seen in the lower left of Fig.~\ref{fig::pdf}, it cannot modulate the large-scale UVB fluctuations. The \texttt{C2} simulation results in a PDF which is strongly peaked around the mean value (or in other words has a high kurtosis), compared to the other models. 

Examining the lower right panel of Fig.~\ref{fig::pdf}, we can summarise the role of the MFP on the UVB distribution. The role of small-scale absorbers is two-fold as they will profoundly affect the small-scale absorptions as well as modulate the MFP of ionising photons and, hence, the large-scale UVB. By modelling the presence of small-scale absorbers, we can reproduce the large-scale UVB fluctuations. However, the PDF converges to a log-normal distribution for the \texttt{pLLS} model, highlighting that the UVB is sensitive to the outliers induced by the small-scale absorbers. Furthermore, the \texttt{C2} model does not produce the large-scale UVB fluctuations to the same extent. In the following sections, we show how these effects on the UVB will also fundamentally impact the observables of reionisation.

\section{Observables of the End of Reionisation}
\label{sec:EoR_Observables}
After having analysed the reionisation histories and the UVB properties of our set of simulations, we will now consider three key observables for the EndEoR, namely the fluctuations in the Ly$\alpha$ opacity, the MFP for ionising photons and the 21-cm power spectrum from the EoR. Measurements for the first two already exist, the 21-cm power spectrum will hopefully be measured in the near future.

\subsection{The Gunn-Peterson optical depth}
\label{subsec:eff_lya_opacity}
A probe for the state of the IGM along the line of sight can be found by analysing the \lya transmission in quasar spectra \citep{Fan2006}. The observed large-scale fluctuations in high-$z$ quasar spectra are an active area of exploration \citep[see][and references within]{Gaikwad2023}. Interpretation of the results coupled with numerical modelling indicates that the fluctuations cannot solely be explained by a uniform UVB in a varying density field \citep[see][and references within]{Ishimoto2022}. Possible explanations of the fluctuations which have been proposed are the effect of density fluctuations in the IGM and temperature fluctuations in the IGM \citep[eg.][]{D'Aloisio2015, Keating2018}. \citet{Davies2016} found that a fluctuating UVB shaped by the presence of self-shielding systems plays a significant role in their models. Including the fluctuating MFP substantially improves agreement with observations. Therefore, modelling the effective \lya optical depth ($\tau_{\mathrm{eff}}$) can not only constrain models of reionisation but can also be used as a probe of the role of small-scale absorbers. 
In this section, we simulate mock observations of the \lya effective optical depth and study how the choice of MFP implementation and the presence of neutral islands impact these observations.

The \lya optical depth can be written as follows \citep[e.g.][]{Pagano2021}:
\begin{equation}
    \label{eq::ly_opt_depth}
    \tau_{\mathrm{Ly}\alpha} = \int_{z_{\mathrm{emi}}}^{z_{\mathrm{obs}}}  n_{\mathrm{HI}} \sigma_{\mathrm{Ly}\alpha}(\lambda, T) \frac{cdt}{dz} dz,
\end{equation}
where $n_{\mathrm{HI}}$ is the column density of neutral hydrogen, $\sigma_{\mathrm{Ly}\alpha}$ is the \lya cross section, which depends on the wavelength $\lambda$ and the temperature of the system $T$. To properly describe the \lya absorption profile and its temperature dependence, we model it with a Voigt profile using the methodology from \citet{Gracia2006}
\begin{equation}
    \label{eq:garcia_cross_section}
    \sigma_{\mathrm{Ly}\alpha}(\lambda, T) = \frac{e^{2} \lambda_{i}^{2} f_{i} \Gamma_{i}}{4 \sqrt{\pi}m_{e}c b^{2}} H(a,x), \quad a = \lambda^{2} \Gamma_{i}/ 4 \pi c b. 
\end{equation}
$\Gamma_i$ and $\lambda_{i}$ are the damping constant and the wavelength of \lya while $b = \sqrt{2kT/m}$ is the Doppler parameter. In this study, we assign the temperature of the ionised cells following the methodology described in \citet{Qin2021} (see eq.8) as follows
\begin{equation}
    T_{g}^{\gamma} = T_{\mathrm{ion,I}}^{\gamma} \bigg[ \bigg( \frac{\mathcal{Z}}{\mathcal{Z_{\mathrm{ion}}}} \bigg)^{3} \frac{\rho_{b}}{\rho_{\mathrm{b, ion}}} \bigg]^{\frac{2 \gamma}{3}} \frac{\mathrm{exp}(\mathcal{Z}^{2.5})}{\mathrm{exp}(\mathcal{Z}^{2.5}_{\mathrm{ion}})} + T_{\mathrm{lim}}^{\gamma} \frac{\rho_{b}}{\overline{\rho_{b}}},
\end{equation}
where $\mathcal{Z} = (1+z)/7.1$ and equation of state index is $\gamma = 1.7$ while $\rho_{b}$ is the baryon density \citep[see][for a detailed analysis]{McQuinn2016}. The subscript 'ion' indicates the quantity at the ionisation redshift $z_{\mathrm{ion}}$, $T_{\mathrm{ion,I}}$ is the temperature after the I-front has propagated, which is assumed to be constant $T_{\mathrm{ion,I}} = 2 \times 10^{4}$ K. The final relaxation temperature is $T_{\mathrm{lim}}^{\gamma} = 1.775 \mathcal{Z} \times 10^{4}$ K.\footnote{Neutral cells are assumed to have a temperature of around 200K.} Additionally, $H(a,x)$ is the Voigt-Hjerting function:
\begin{equation}
    \label{eq::hjerting_func}
    H(a,x) = \frac{a}{\pi} \int_{- \infty}^{\infty} dy \frac{\mathrm{exp}[-y^{2}]}{(x-y)^{2}+a^{2}}, \quad \mathrm{where} \quad  y = v/b.
\end{equation}
We measure the transmitted flux in our simulations averaged over $N$ cells. See sec.~5 of \citetalias{Bosman2022} for more details. The effective optical depth\footnote{\label{foornote:Lya}Note that we calculate the \lya optical depth from the neutral hydrogen on the grid and do not include the effect of the subgrid neutral hydrogen assumed to be present and absorbing ionising photons in the \texttt{LLS40, eLLS and pLLS} simulations. This allows us to make a direct comparison with the other two simulations, \texttt{r40} and \texttt{C2}, which do not introduce subgrid \HI. However, it does mean we may be underestimating the actual \lya opacity implied by these models. This may explain why the mean \lya opacity at a fixed neutral fraction (but different redshifts) in Tab.~\ref{tab:PDF_means} is lower for the diffuse models when compared to \texttt{C2}, though note $\tau_{\mathrm{eff}}$ is also redshift-dependent.} is given as
\begin{equation}
    \label{eq::eff_lya_opt_depth}
    \tau_{\mathrm{eff}} = - \ln \bigg( \frac{1}{N} \sum_{i} \mathrm{exp}\big[-\tau_{\mathrm{Ly}\alpha}(i) \big]  \bigg),
\end{equation}
Our results are computed for transmission over 30 cMpc skewers, to avoid the evolution of the signal along the LOS in the observable, while we also generate 70 cMpc skewers to calculate the fit parameters from Tab.~\ref{tab:PDF_means} to match the fit values from \citetalias{Bosman2022}. We sample $5 \times 10^{3}$ rays for each simulation in this study, taking a random direction on the $x$,$y$, and $z$ planes to construct the skewer. Note we have also conducted checks with different sample sizes as well as sampling the rays at random locations versus at the position of the ionising sources. In both cases, we find our results to be consistent.

\subsubsection{Cumulative Distribution Functions of the effective Lyman-$\alpha$ opacity (CDFs)}
\begin{figure*}
    \includegraphics[width=.49\linewidth]{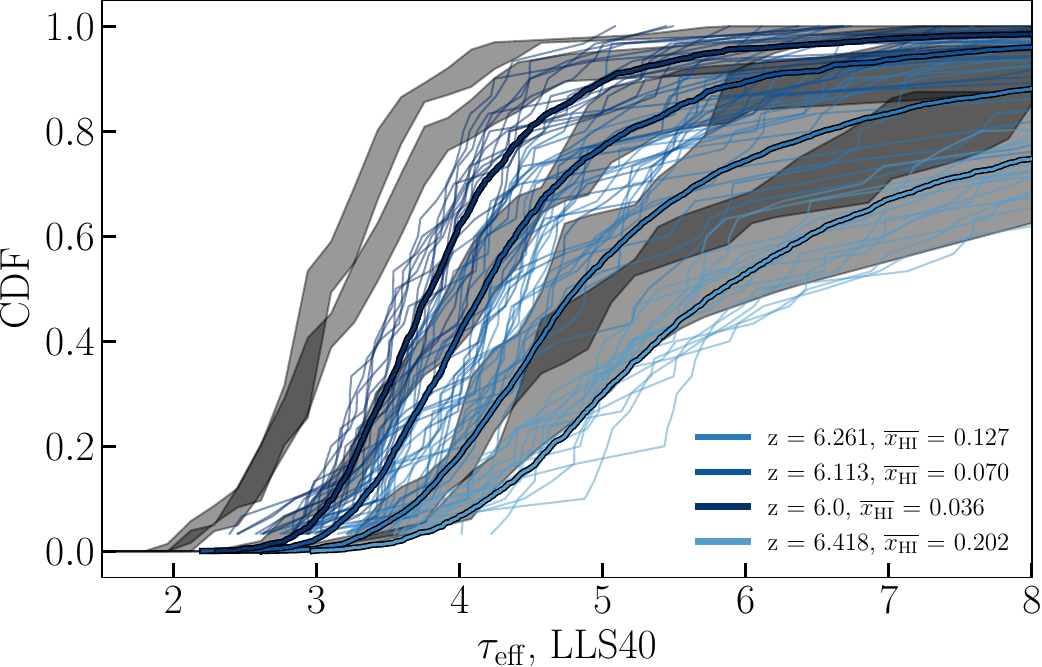}
    \includegraphics[width=.49\linewidth]{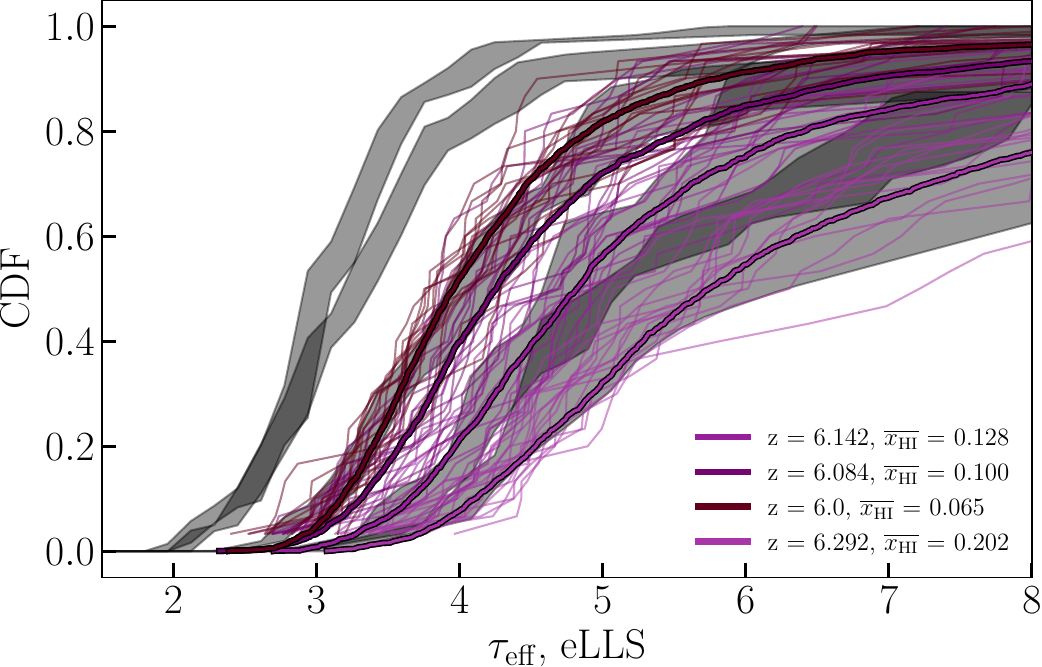}
    \includegraphics[width=.49\linewidth]{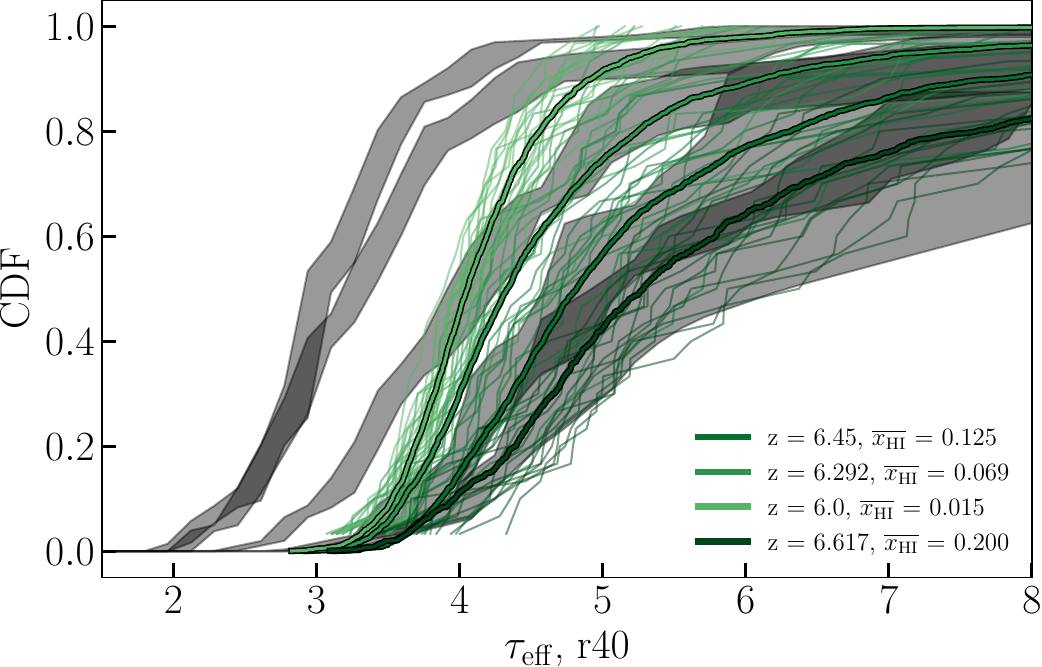}
    \includegraphics[width=.49\linewidth]{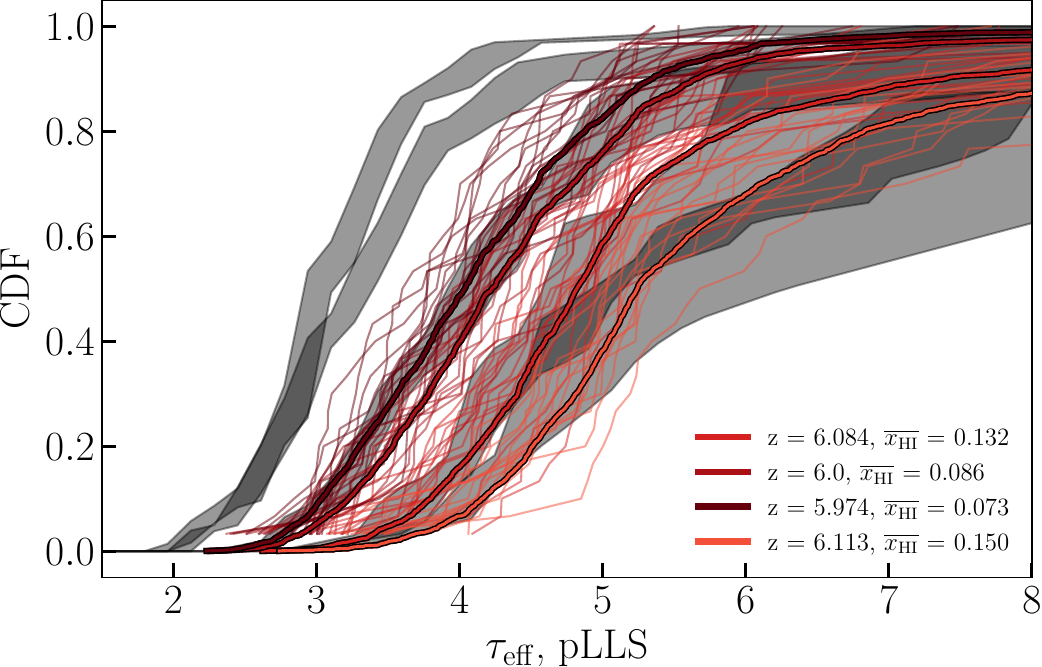}
    \includegraphics[width=.49\linewidth]{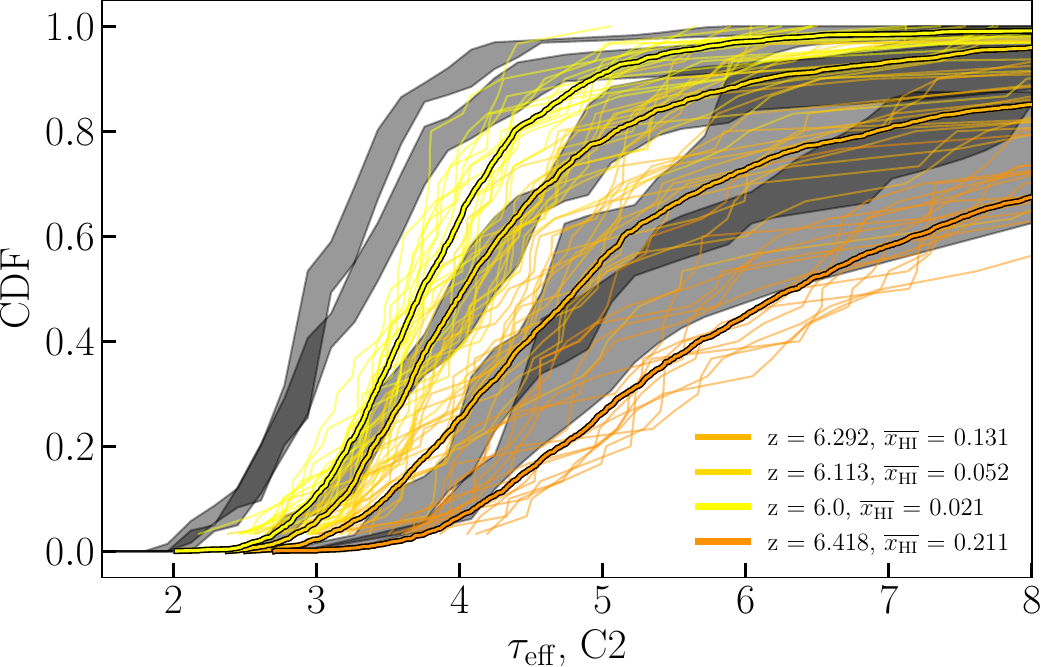}
    \includegraphics[width=.49\linewidth]{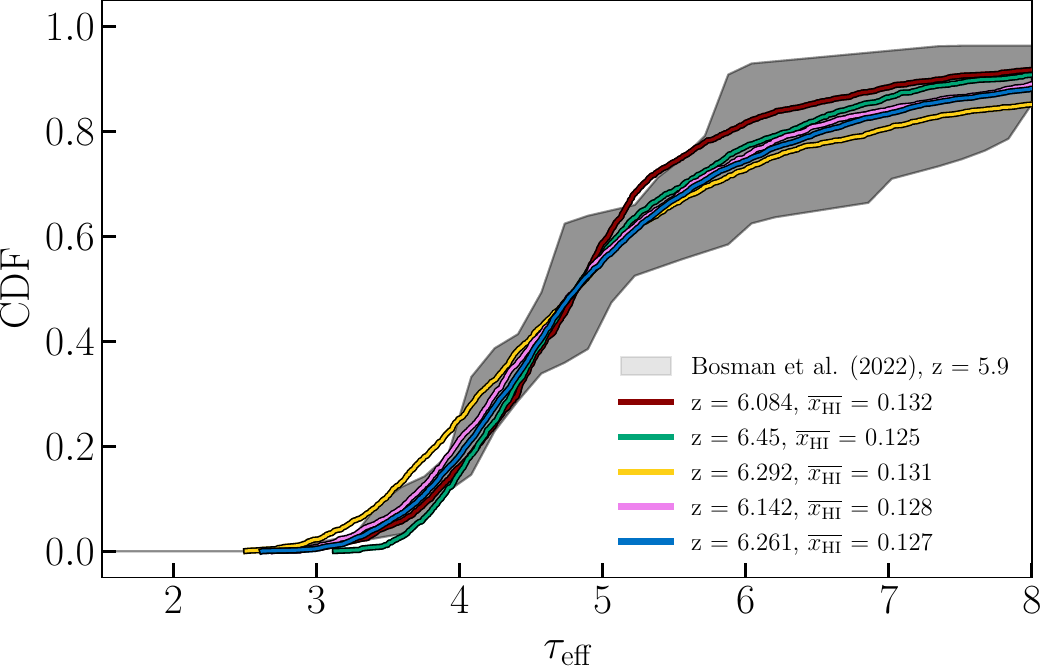}
    \caption{The Cumulative Distribution Functions (CDFs) of the effective \lya opacity across 30 cMpc during the final stages of reionisation. The models and their nomenclature are described in Tab.~\ref{tab:C2sims}. The dark shaded regions are the 1$\sigma$ uncertainties of the measurements reported by \citetalias{Bosman2022} (for $z=  5.525, 5.620, 5.716,  5.915, 6.017$), in order to follow the EndEoR as estimated in tab.~3 of \citet{Gaikwad2023}. The CDFs for each model are compared at a fixed neutral fraction and mean \lya forest transmission, such that the median of the transmission matches that of the data at $z \approx 5.9$. The skewing of each CDF with increasing redshift is indicative of lines of sight with a higher \lya opacity, hinting that the EoR is still underway. All simulated skewers chosen within this analysis are averaged over 30 cMpc, to match the reported data. While each thick line is the mean distribution of $\tau_{\mathrm{eff}}$, the 10 thin lines represent random LOS, which have been chosen to highlight the variation of the distribution. The bottom right panel shows the simulated and real data at redshift 5.9, normalised against the mean value of each distribution. Including redshift and position dependence in the LLS implementation (\texttt{pLLS} model) leads to a more accurate model of observations. The \texttt{r40} model (left middle row in green) is quantitatively excluded as it fails to reproduce high effective \lya opacity at the end of reionisation. Meanwhile, models such as  \texttt{eLLS} and  \texttt{C2} (right top row in purple and left bottom row in yellow, respectively) can marginally reproduce the observables.}
    \label{fig2}
\end{figure*}
\begin{figure*}
    \includegraphics[width=.49\linewidth]{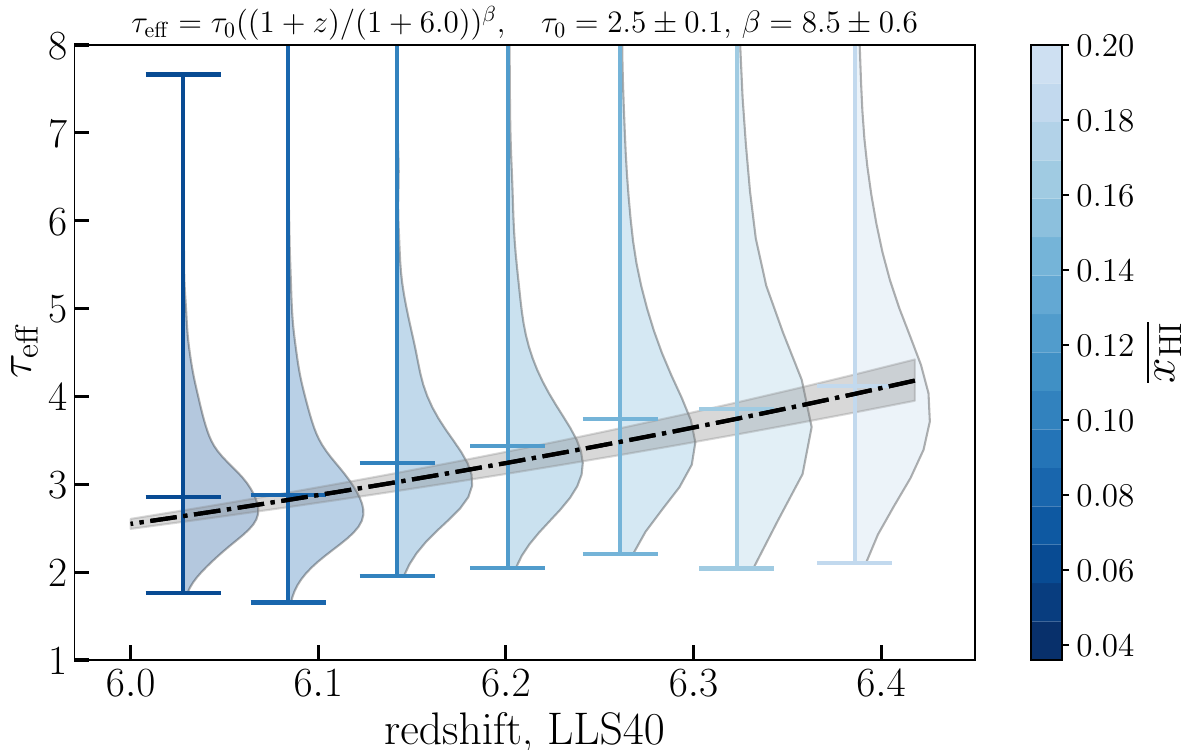}
    \includegraphics[width=.49\linewidth]{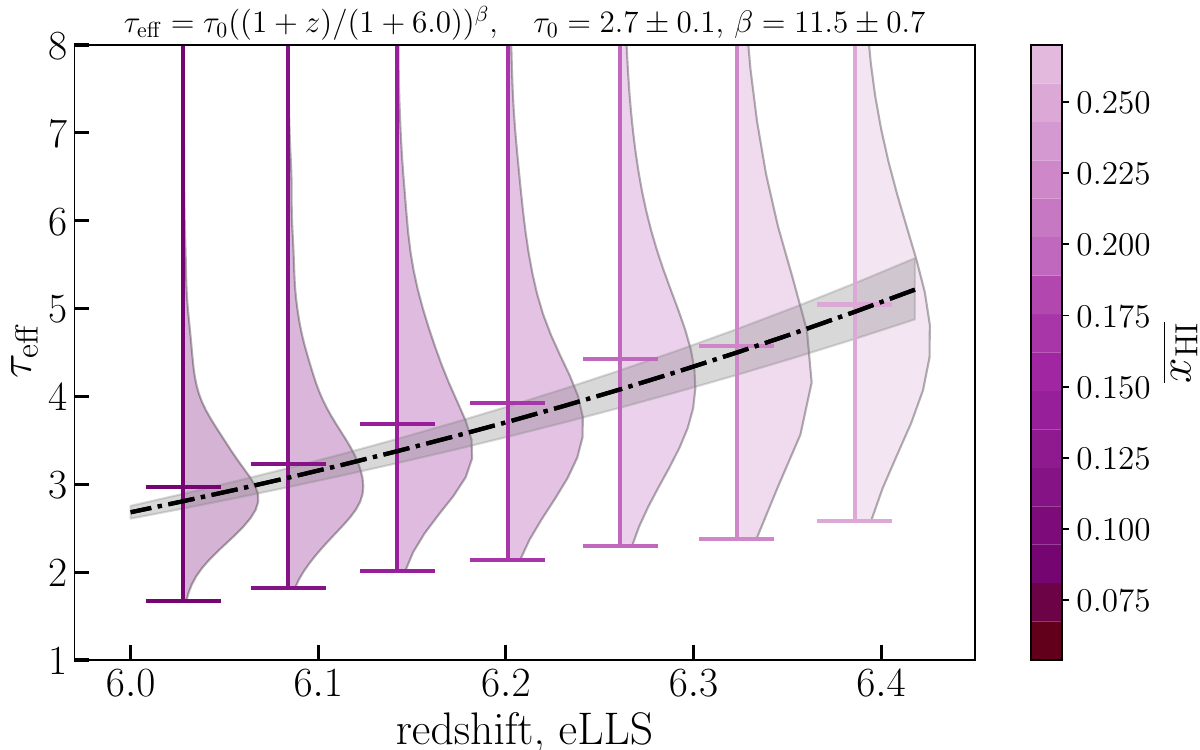}
    \includegraphics[width=.49\linewidth]{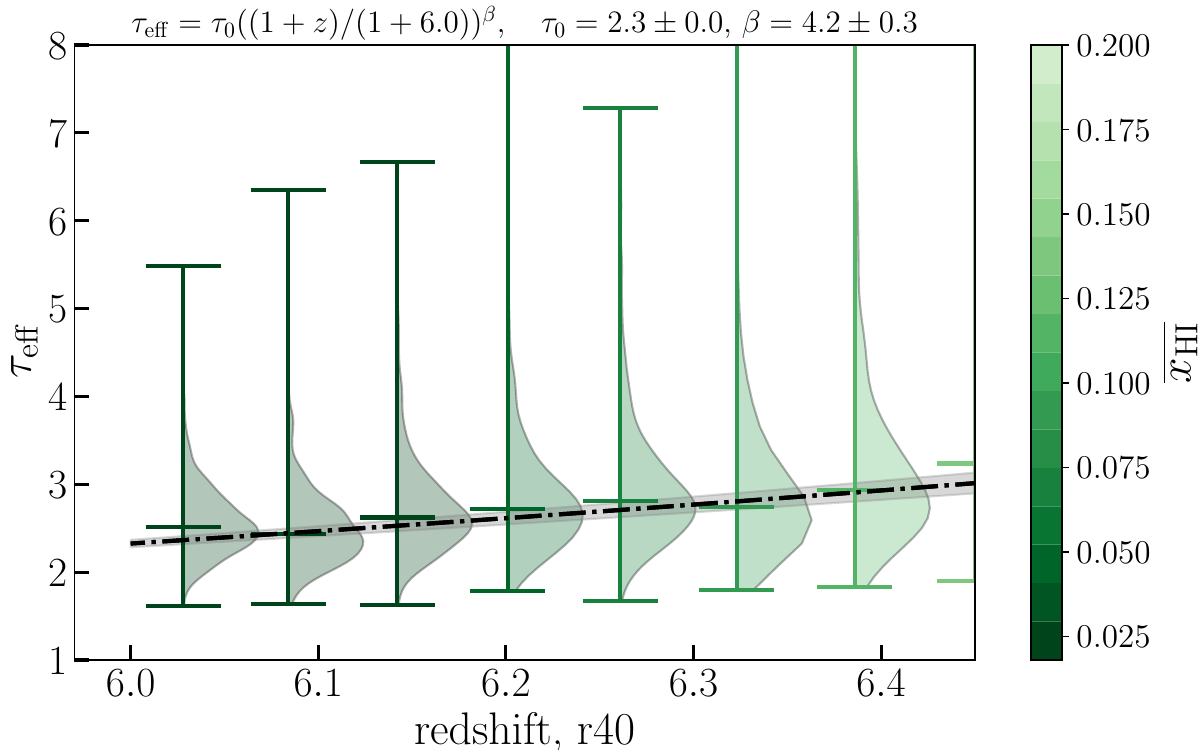}
    \includegraphics[width=.49\linewidth]{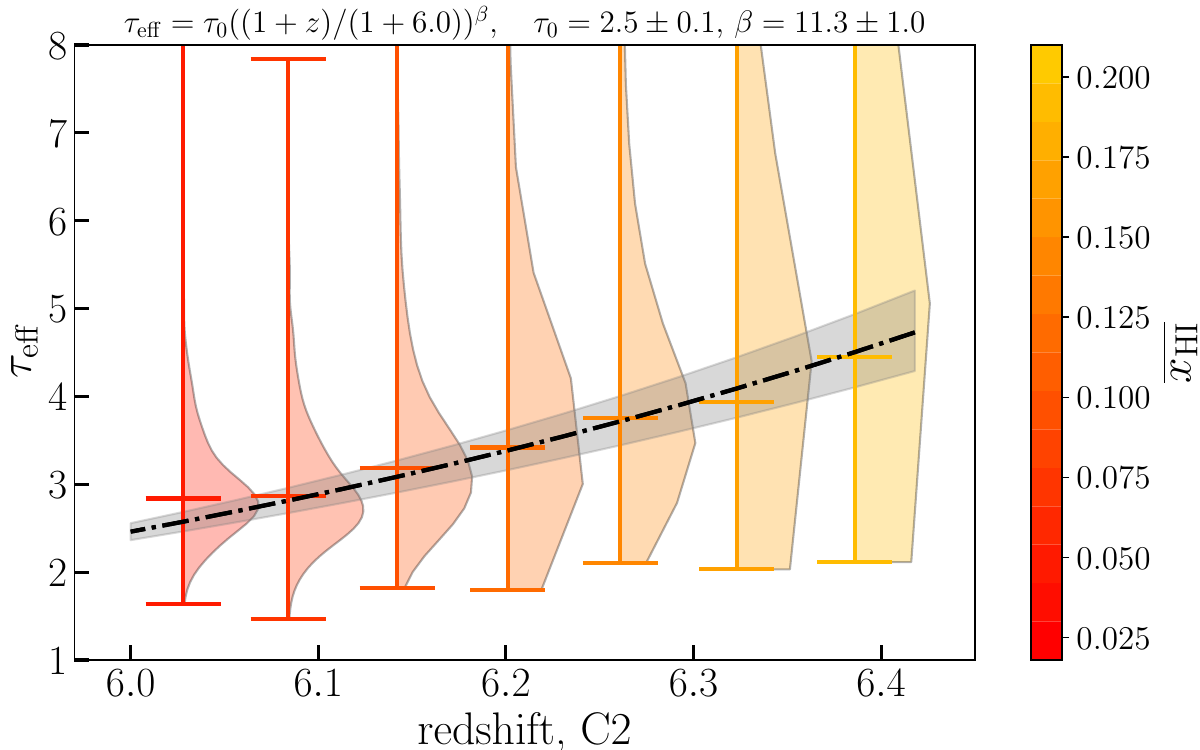}    \includegraphics[width=.49\linewidth]{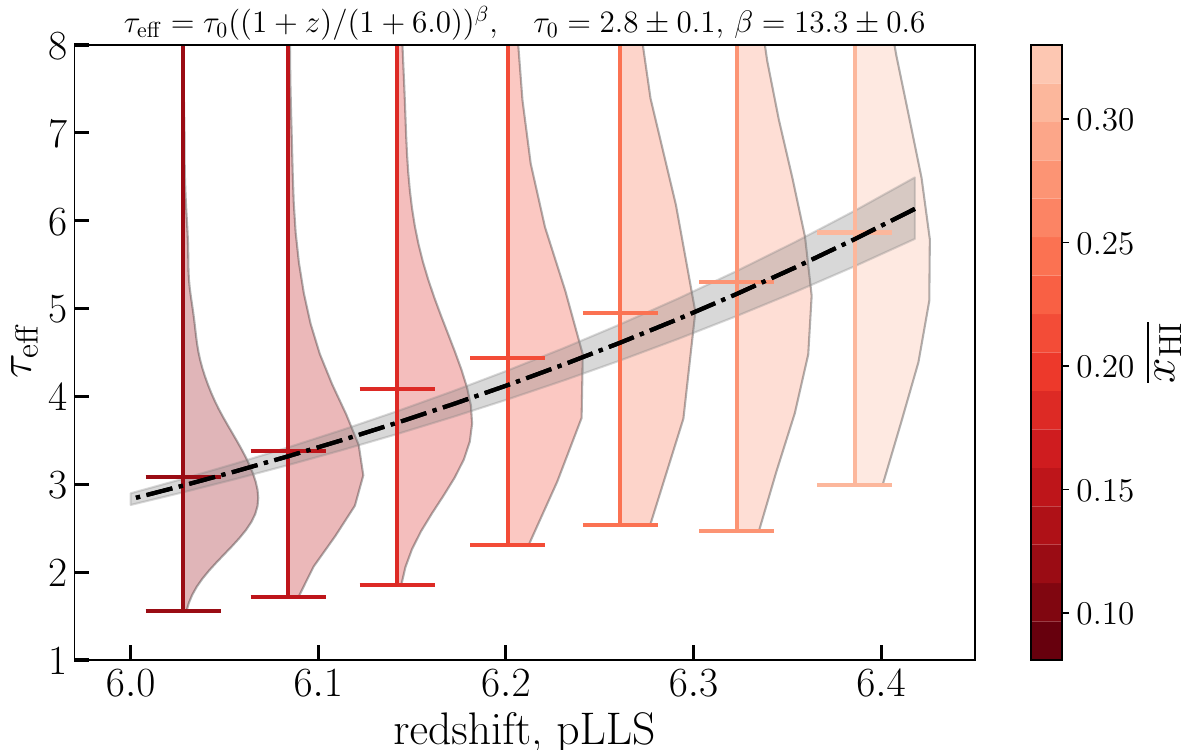}
    \caption{Violin plots illustrating the evolution of the distribution of the \lya opacity across 70 cMpc skewers during the final stages of reionisation. The black dashed lines represent the model of $\tau_{\mathrm{eff}}$ in equation~\ref{eq:tau_eff_fit}, given the best-fit parameters and their respective uncertainties from Tab.~\ref{tab:PDF_means} compared to the values reported from \citetalias{Bosman2022}. Though the shape of the widths in each violin plot are less informative, this visualisation of the data highlights the differences of shape and spread of the \lya optical depth for each model during the EnDoER and is complimentary to Fig.~\ref{fig2}. }
    \label{fig5}
\end{figure*}
We present the Cumulative Distribution Function (CDF) for our set of models against the data reported by \citetalias{Bosman2022} (seen in grey) in Fig.~\ref{fig2}. We focus our discussion on the final stages of reionisation, visualising output for redshifts where the global neutral fraction is $\xb \leq 0.20$, for the first three rows of Fig.~\ref{fig2} (thick lines). To highlight the effect of sample variance in our analysis we visualise randomly selected CDFs (thin lines), choosing 10 for clarity. The analysis in \citet{Gaikwad2023} reports a neutral fraction of $\xb \approx 0.128$ at a redshift 5.9. Hence, we compare our models to the data at a fixed neutral fraction and mean \lya forest transition. To achieve this, we rescale the mean transition to that measured by \citetalias{Bosman2022} at $z = 5.9$, where the quasar sample is robust. Therefore, we can study the shape of the functions to gain insight into the distributions and study the different implementations of the effects of small-scale absorbers. Hence, in the right panel of the third row of Figure~\ref{fig2}, we include each output of our simulation closest to the observed CDF, all of which are normalised against their mean value.

Lastly, we also perform a best-fit analysis of the data following the parametrisation\footnote{The original form of the equation includes a third parameter $C$, which we do not consider as we equate $\tau_{\mathrm{eff}} (1+z_0) = \tau_0$. The $C$ parameter is more relevant for data with implicit noise, which is not included in our simulations.} of $\tau_{\mathrm{eff}}$ reported in eq.3 of \citetalias{Bosman2022} that is
\begin{equation}
\label{eq:tau_eff_fit}
    \tau_{\mathrm{eff}} (1+z) = \tau_0 \bigg( \frac{1+z}{1+z_0} \bigg)^{\beta}, 
\end{equation}
where $z$ and $\beta$ are free parameters, and $z_{0}=6$. Our best-fit values are listed in Tab.~\ref{tab:PDF_means} as well as the mean $\tau_{\mathrm{eff}}$ predicted by the fit for each model, at the redshift where $\xb \approx 0.128$ (also see Fig.~\ref{fig5}). The CDF results of \citetalias{Bosman2022} at redshifts of 5 rise rapidly from zero to unity, indicative of Gaussian distribution, and become slanted with increasing redshift due to the positive skewness of distribution attributed to the presence of neutral hydrogen. Each $\tau_{\mathrm{eff}}$ is the cumulative value of a spatial skewer, which itself captures diverse environments of different density, temperature, and ionisation states. Hence, the effective value will capture some average quantity of the IGM across the line of sight. How the peak of the distribution evolves is therefore related to both the evolution of the UVB as well as the disappearance of \HI islands. For example, at high-$z$, the 1$\sigma$ distribution reported by \citetalias{Bosman2022} (rightmost grey CDFs of Fig.~\ref{fig2}) are positively skewed, hinting at an IGM with a higher fraction of residual neutral hydrogen as well as the existence of neutral islands.

The mean of the distribution is shifted due to the evolving residual neutral fraction, while the presence of neutral islands results in a higher $\tau_{\mathrm{eff}}$, which positively skews the data. The \texttt{r40} model (in green, middle row, left column) does not exhibit the described skewness, with a visibly slight variance from sight-line to sight-line. Following Section~\ref{sec::UVB_PDF_analysis}, the implementation of the cut-off barrier results in ionised regions with a high-amplitude, ubiquitous, and slowly evolving UVB value across redshift. The fluctuations in the UVB, and hence the \lya optical depth, are then set either by the density field, the 40 cMpc barrier to ionising photons, or the neutral islands. The slow evolution of the UVB, noted in Fig.~\ref{fig:1}, results in a weaker redshift revolution of $\tau_{\mathrm{eff}}$ seen in the low value of $\beta = 4.2 \pm 0.3$ Tab.~\ref{tab:PDF_means}, compared to the observations (where $\beta = 13.7 \pm 1.5$). Moreover, the relatively small scatter in the random sight lines is due to the suppression of the fluctuations in the IGM, which is, on average, more irradiated compared to the other models.

Conversely, the \texttt{pLLS} model is marked by a rapid evolution of the CDF against redshift, which is consistent with the values reported by \citetalias{Bosman2022}. Although the simulation has not completely reionised in Fig.~\ref{fig2} (middle row, right column, in red), we can observe this rapid evolution in tandem with the positive skewness of the distribution measured in observations. As discussed in Section~\ref{sec::UVB_PDF_analysis}, introducing the effect of small-scale absorbers results in fluctuations of the UVB in dense regions. Lastly, the rapid evolution of the lower values of $\tau_{\mathrm{eff}}$ seen in the \texttt{pLLS} model is more consistent with those observed in the data shown in grey. 

The \texttt{eLLS} and \texttt{LLS40} simulations exhibit similar characteristics to the \texttt{pLLS} model. However, the evolution of the CDFs is impeded by excluding the density-dependence of the LLS (which results in small-scale fluctuations in the UVB for \texttt{pLLS} as discussed in Section~\ref{sec::UVB_PDF_analysis}), resulting in a slower evolution of the low values of $\tau_{\mathrm{eff}}$. Nevertheless, \texttt{eLLS} is broadly consistent in shape and evolution with the global features of the \lya CDFs, as evidenced by the redshift evolution of $\beta = 11.5 \pm 0.7$, closest of the remaining models to the value derived from observations. 

Lastly, the clumping simulation \texttt{C2} reproduces the broad features of the measurements for high \lya opacities, similar to the data. The evolution of the mean value is slower than those noted in observations ($\beta = 11.3 \pm 1.3$), and the evolution of the lowest values of $\tau_{\mathrm{eff}}$ appears to be impeded. 

Our conclusion here follows the one from Section~\ref{sec::UVB_PDF_analysis}; the model in which the effect of small-scale absorbers depends on both redshift and position reproduces the observations better. The cut-off barrier model fails to reproduce the high effective \lya opacity across EndEoR. Models such as \texttt{eLLS} and \texttt{C2} can marginally reproduce the observables while \texttt{pLLS} is most successful in reproducing the features inferred from the observational data. 
\begin{table}
    \centering
    \caption{Best fit parameters of equation~\eqref{eq:tau_eff_fit} of the effective \lya opacities ($\tau_{\mathrm{eff}}$), derived for 70 cMpc skewers to match the reported values from \citetalias{Bosman2022}. Note that we choose $z_0 = 6$ for our fit, which is our lowest available redshift. In the third column the mean $\tau_{\mathrm{eff}}$ predicted by the fit for each model, at the redshift where $\xb \approx 0.128$. All models, save for \texttt{r40}, roughly agree with the observations of \citetalias{Bosman2022} (see Footnote~\ref{foornote:Lya}), given the estimate of the global neutral fraction from \citet{Gaikwad2023}.}
    \renewcommand{\arraystretch}{1.2} 
    \begin{tabular}{|p{3.5cm}|c|c|c|}
        \hline
        Label & $\tau_0$ & $\beta$ & $ \tau_{\mathrm{eff}}$ at \\
         &  &  & $\xb \approx 0.128$ \\
        
        \hline \hline
        \citetalias{Bosman2022} & 0.3 $\pm$ 0.1 & 13.7 $\pm$ 1.5 & 4.6 $\pm$ 1.4\\
        \hline
        \texttt{r40}, cut-off fixed barrier & 2.3 $\pm$ 0.1 & 4.2 $\pm$ 0.3 & 3.0 $\pm$ 0.1 \\
        \hline
        \texttt{C2}, global clumping parameter of two & 2.5 $\pm$ 0.1 & 11.3 $\pm$ 1.0 & 4.0 $\pm$ 0.2 \\
        \hline
        \texttt{LLS40}, diffuse fixed barrier & 2.5 $\pm$ 0.1 & 8.5 $\pm$ 0.6 & 3.5 $\pm$ 0.2 \\
        \hline
        \texttt{eLLS}, diffuse evolving barrier & 2.7 $\pm$ 0.1 & 11.5 $\pm$ 0.7 & 3.4 $\pm$ 0.1 \\
        \hline
        \texttt{pLLS}, density-dependent diffuse barrier & 2.8 $\pm$ 0.1 & 13.3 $\pm$ 0.6 & 3.3 $\pm$ 0.1 \\
        \hline
    \end{tabular}
    \label{tab:PDF_means}
\end{table}
\subsection{The mean free path of ionising photons from the EoR.}
\label{Sec:MFP_Study}
\begin{figure}
\includegraphics[width=\columnwidth]{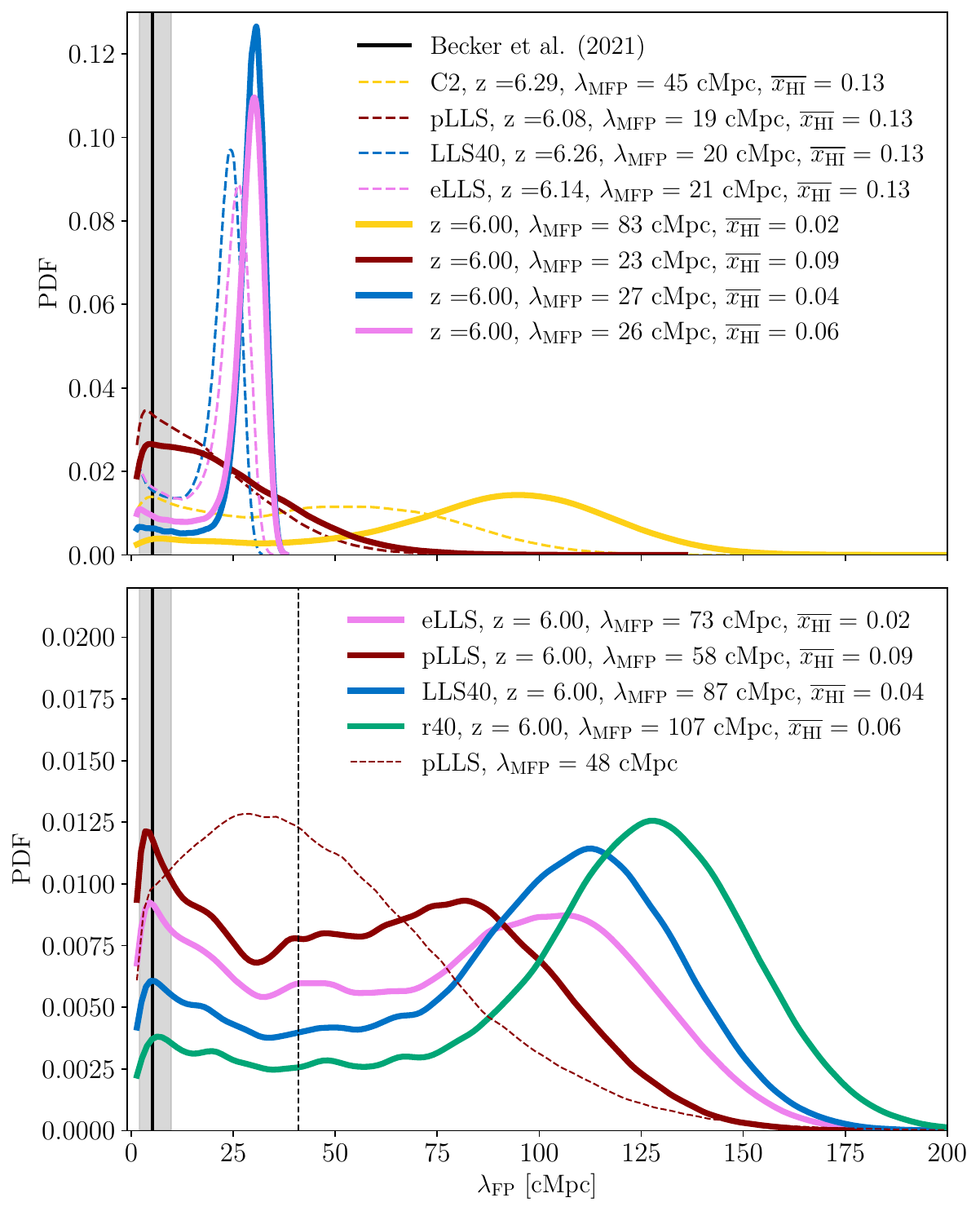}
\caption{Upper Panel: PDF distribution of the free path of ionising photons within the simulations reported in Tab.~\ref{tab:C2sims}. The thick solid lines are the PDF functions at redshift 6, while the thin dashed lines are normalised to a fixed neutral fraction of 13~\%. The colour scheme and naming notation are reported across this body of work while the thick solid black line and its associated gray area is the measurement of the MFP by \citet{Becker2021} at redshift 6. 
Bottom Panel: Distribution for the cut-off barrier model \texttt{r40} as well as the PDFs of the free paths of the diffuse barrier models, excluding the role of the additional opacity added to model the LLS. Conversely, the thin red line is the same analysis conducted only on the small-scale absorbers used in the \texttt{pLLS} model, and is consistent with the thin black line is the reported MFP value by \citetalias{Worseck2014}. On the one hand, the \texttt{r40} and \texttt{C2} models cannot reproduce the small values of the free paths reported in \citet{Becker2021,Bosman2021,Gaikwad2023}. On the other hand, the diffuse barrier models, \texttt{LLS40} and \texttt{eLLS}, can match the MFP. It is only the \texttt{pLLS} model seen at the top panel in red, which reproduces a PDF of the MFP, which both matches the mean value and has a large variation, favouring small free paths.}
\label{fig:MFP}
\end{figure}

We recreate the ray-tracing approach described in Section~\ref{subsec:eff_lya_opacity} and follow the Lyman continuum photons in the ionised medium during the EndEoR for a given LOS. We calculate the opacity to Lyman continuum radiation as 
\begin{equation}
    \label{eq:MFP}
    \tau_{\mathrm{LyC}} = \int_0^1 n_{\HI} \sigma_{\HI} dx,
\end{equation}
where $n_{\HI}$ and $\sigma_{\HI}$ are the number density of hydrogen (see equation~\eqref{eq::UVB_Becker21}). Following the definition of the MFP of ionising photons, we trace the cumulative opacity to ionising radiation up to a value of unity ($\tau_{\mathrm{LyC}} \approx 1$) from the positions of the ionising sources in our simulation. This allows us to probe multiple free paths (FP), which probe the IGM at the end of reionisation, exploring the role of residual neutral hydrogen along the lines-of-sight (LOS) and the remaining extended neutral islands. For the sake of consistency, we use exactly the same $2.5 \times 10^{6}$ rays, initiated at the positions of the ionising sources, when analysing each of our five simulations. This way we can better contextualise the direct effect that the implementation of the effect of small-scale absorbers plays in determining the MFP measurements. Additionally, because the diffuse barrier approach model the effect as an additional subgrid component within our simulation (see Section~\ref{subsec:global_params}), simulations \texttt{LLS40},  \texttt{eLLS} and \texttt{pLLS} allow us to perform our analysis either by removing or including the contribution of the subgrid small-scale absorbers. While the absorbers have a time-dependent effect on the opacity in the IGM, performing this analysis allows us to directly probe how the LLS model can affect the MFP observable.

The resulting PDFs are presented in the upper panel of Fig.~\ref{fig:MFP}, for a fixed redshift of 6 (solid) or a fixed neutral fraction of approximately 13~\% (dashed). The bottom panel includes the same analysis of the diffuse barrier simulations excluding the subgrid opacity from the small-scale absorbers. 

What is immediately apparent for the free path PDFs is that the clumping simulation \texttt{C2} at redshift 6 produces a distribution, which is Gaussian and has a mean of around 80 cMpc ($\approx 11$ pMpc). Initially, it would be tempting to relate the large value of the MFP to the low neutral fraction of the simulation ($\approx 2$~\%), however, the PDF for the same model at a higher redshift and neutral reaction (yellow dashed line), remains biased towards large free paths in contrast to current measurements \citep[e.g.][]{Becker2021}. Rather, the behaviour of the distribution is related to the residual neutral hydrogen in the IGM. The additional opacity to the IGM, indirectly introduced through the boosted recombination rate, is insufficient in reducing the length of the free paths of the ionising photons \citep[also see][where much higher clumping values than here are used]{Mao2020}. Because $R_{\mathrm{max}} =70$ cMpc, the majority of ionising sources during EndEoR are free to interact with each other, further ionising the IGM between them and boosting the value of the MFP. 

A similar distribution is also observed for the \texttt{r40} model seen in green in the bottom panel of Fig.~\ref{fig:MFP}. Limiting the propagation of the ionising photons to 40 cMpc for \texttt{r40} when conducting the radiative transfer does not directly reproduce lower free path values when sampled in this manner. The cut-off barrier, which serves as an upper limit of ionising photons, is ineffective at the end of EoR, during which the majority of the cells in the simulation have active sources of radiation, resulting in a highly ionised IGM. Note that the shape of this PDF and most distributions are skewed towards low free path values. We can intuitively imagine the overall PDF of the free paths as a convolution of two different distributions. On the one hand, for small values of the free paths, the function follows a Poisson distribution, sensitive to the rays absorbed by neutral islands in the vicinity of the ionising source. On the other hand, for large values of the free paths of ionising photons, the distribution resembles a normal distribution, as it probes the ionised intergalactic medium, which is typically in a state of equilibrium during the final stages of reionisation.

The same result is observed to a varying degree for the diffuse barrier models \texttt{LLS40, eLLS, pLLS} in the bottom panel of Fig.~\ref{fig:MFP} (solid lines), for the case when the additional subgrid absorption due to small-scale absorbers is removed from the analysis. Despite this, we can see that by including redshift and density dependence in the model of the LLS, the resulting distribution (even with the aforementioned absorbers being removed) converges to distributions skewed to lower values and a lower MFP value. We can, therefore, infer that by including the additional opacity from small-scale absorbers, we not only remove the excess photons which should have interacted with the self-shielded systems but also regulate the overall state of the ionised IGM. 

As an additional result, we can sample the free paths between the density-dependent absorbers included in the \texttt{pLLS} simulation in seen at the bottom panel of Fig.~\ref{fig:MFP} as a dashed red line. We confirm that the produced distribution peaks at a $\lambda_{\mathrm{MFP}} = 48$ cMpc as it has been normalised to match the fit from \citetalias{Worseck2014}. In the case of the additional opacity for the \texttt{LLS40} and \texttt{eLLS}, the additional opacity would peak as a delta function (as $\tau = 1$ at $\lambda_{\mathrm{MFP}} =40$ cMpc by definition). 

It is only when we convolve both of the red functions (dashed and solid) for the \texttt{pLLS} model from the bottom panel of Fig.~\ref{fig:MFP}, that we arrive at the true distribution in the top panel of the same figure (similarly for \texttt{LLS40} and \texttt{eLLS}). Unlike the other diffuse barrier models, whose distributions become shifted and centred at the predefined MFP value, the position-dependent model produces a log-normal distribution skewed towards smaller free paths for ionising photons. At the same time, some rays can travel to as much as 60 cMpc. This model is more natural than \texttt{LLS40} and \texttt{eLLS}, both of which, due to the scaling, disfavour low and high values of the free paths alike. The complete PDF produced by the \texttt{pLLS} run naturally reproduces a large variety of possible sight-lines such as those reported \citet{Bosman2021} (see fig.~2) as well as \citet{Rahmati2018}, some more consistent with the ionised universe, while others are more sensitive to the presence of small-scale absorbers or neutral islands. Crucially, this is confirmed when examining the distributions for a fixed neutral fraction (dashed lines in the top panel), highlighting the conclusion that a position-dependent model of the LLS affects both the small and large-scale physics of the IGM. This is in concordance with the UVB results in sec.~\ref{sec::UVB_PDF_analysis}.
\begin{figure}
\includegraphics[width=\columnwidth]{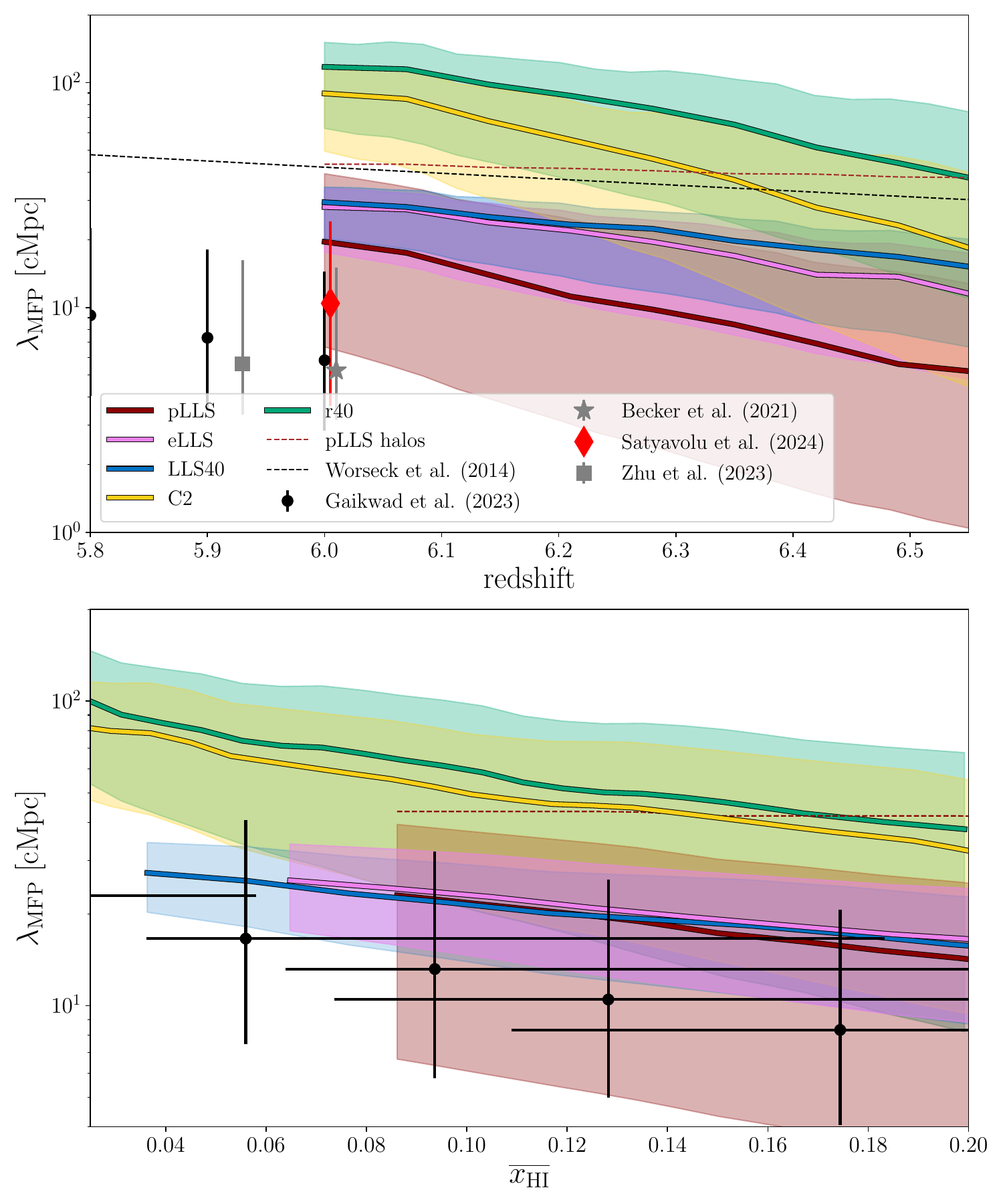}
\caption{Evolution of the median value of the free path evolution over the end stages of reionisation against redshift (top panel) and the volume-averaged neutral fraction (bottom panel) as well as the standard deviation around the mean value (shaded areas). The labels and colouring scheme follow that in Fig.~\ref{fig:MFP}. The black dashed line is the MFP evolution depicted in \citetalias{Worseck2014}, whilst the red dashed line is the MFP of the LLS model implemented for the \texttt{pLLS} simulation. Measurements of the MFP from \citet{Becker2021}, its re-analysis from \citet{Satyavolu2024}, and \citet{Zhu2023} are presented in the upper panel whilst the estimates of the MFP from \citet{Gaikwad2023} are presented in both panels. It is only the free path distributions from the diffuse barrier models (\texttt{LLS40, eLLS,pLLS}) in blue, fuchsia, and red, respectively, which agree with the values from \citet{Gaikwad2023}.}
\label{fig:MFP_evolve}
\end{figure}

Lastly, in Fig.~\ref{fig:MFP_evolve}, we plot the evolution of the mean free path of ionising photons for each of our simulations, against redshift and the global neutral fraction (upper and lower panels, respectively). We include the MFP observations by \citet{Becker2021}, \citet{Zhu2023}, and \citet{Satyavolu2024} as well as the estimates from \citet{Gaikwad2023}. Moreover, we choose to represent the evolution of the median value of each distribution but also include the $1\sigma$ around the mean value to better capture the full range of the statistics. In addition, we compare the MFP evolution from \citetalias{Worseck2014} (from equation~\eqref{eq:W14fit}, seen in black) to that derived from our small-scale absorbers within the simulation. The inferred evolution slightly differs in amplitude and is biased towards larger values of the MFP. This can be attributed to the normalisation, which assumes that the global opacity must be normalised against the reported mean free path. As expected, the mean values derived for each of the diffuse models are significantly lower than for the remaining models. Most notably, the reported measurements of the MFP from \citet{Bosman2021} and the estimates from \citet{Gaikwad2023} are consistent within the distribution of free paths from the \texttt{LLS40}, \texttt{eLLS}, and \texttt{pLLS} models when compared at a fixed neutral fraction. However, examining the full PDF of free paths for the diffuse models (dashed lines) in the upper panel in Fig.~\ref{fig:MFP}, we see that only the peak and spread of the \texttt{pLLS} model are consistent with the measurement from \citet{Becker2021}), highlighting the importance of the spatial distribution of small-scale absorbers in modulating the MFP of ionising photons.
\subsection{The 21-cm signal from the epoch of reionisation}
\label{sec:21cm_ps}
In this section, we turn our attention to the 21-cm power spectrum from reionisation. Compared to the observables discussed in Sec.~\ref{eq::eff_lya_opt_depth} and \ref{Sec:MFP_Study}, which probe the ionised IGM, the 21-cm signal is sensitive to the neutral hydrogen present in the IGM. We examine the impact LLS modelling has on the shape and amplitude of power spectrum.
\subsubsection{Differential brightness temperature}

The 21-cm line is the result of the spin-flip transition, occurring in the ground state of neutral hydrogen. The 21-cm radiation observed by radio interferometry telescopes is defined as the differential surface temperature against the CMB. In the Rayleigh-Jeans limit it can be written as follows \citep[e.g.][]{Mellema2013}
\begin{equation}
\label{eq::Tb}
\delta T_{\mathrm{b}} (\mathbf{r}, z) =\frac{1-e^{-\tau(\mathbf{r}, z)}}{1+z}(T_{s}(\mathbf{r}, z)-T_{\mathrm{CMB}}(z))\,,
\end{equation}
where $\tau$ is the optical depth of the gas and $T_{s}$ is the spin temperature. 
The spin temperature characterises the level populations of the 21-cm transition and has to be different from $T_\mathrm{CMB}$ to produce an observable signal. At the redshifts of reionisation $T_{s}$ is expected to be similar to the kinetic temperature of the gas due to the frequent interaction of the intergalactic $\HI$ with \lya photons produced by star-forming galaxies \citep{Field1959, Wouthuysen1952}. X-ray heating caused by the same galaxies is expected to increase this kinetic temperature \citep{Pritchard2007}. 

Assuming an optically thin gas $\tau \ll 1$ at high redshifts, equation~\ref{eq::Tb} can be rewritten as 
\begin{align}
\label{eq::TbS}
\delta T_{\mathrm{b}} (\mathbf{r}, z) = T_{0} (\mathbf{r},z) \xb(1+\delta_{\HI}(\mathbf{r}))(1+\delta_{\rho}(\mathbf{r})) \ , 
\end{align}
\begin{align}
&T_0 (\mathbf{r},z) \approx 27 
\bigg(\frac{1+z}{10}\bigg)^{1/2} \bigg(\frac{T_{\mathrm{s}}(\mathbf{r},z) -T_{\mathrm{CMB}}(z) }{T_{\mathrm{s}}(\mathbf{r},z) }\bigg) 
\nonumber
\\ &\bigg( \frac{\Omega_{\mathrm{b}}}{0.044}  \frac{h}{0.7} \bigg) \bigg( \frac{\Omega_{\mathrm{m}}}{0.27}  \bigg)^{-1/2} 
\bigg( \frac{1-Y_{\mathrm{p}}}{1-0.248} \bigg)
~\mathrm{mK},  
\end{align}
Here the variable $\rho$ is the mass density, whilst $\delta_{\rho}= \rho/\bar{\rho} -1 $ is the corresponding fluctuation. Similarly, $\delta_{\HI}$ denotes the fluctuation of the neutral fraction $\xHI$ field. The term $T_{0}$ collects all the cosmological parameters as well as the spin temperature dependence. Note that within this work we do not consider the effects of redshift space distortions.
We analyse the 1D spherically averaged power spectra of the 21 cm signal shown in equation~\ref{eq::TbS}. We also define the dimensionless power spectrum at a certain wave-number $k$ as $\Delta^{2}(k)=k^{3} P(k)/(2\pi^{2})$.

A large variety of upper limit values on the 21\,cm power spectrum have been reported by several low-frequency radio interferometers (at various redshifts and scales), e.g., the Low-Frequency Array (LOFAR)\footnote{\url{www.lofar.org}} \citep{Mertens2020}, the Hydrogen Epoch of Reionization Array (HERA)\footnote{\url{https://reionization.org}} \citep{HERA2021, TheHERACollaboration_2022}, the Murchison Widefield Array (MWA)\footnote{\url{www.mwatelescope.org}} \citep{Trott2020,Yoshiura2021}, the Giant Metrewave Radio Telescope (GMRT)\footnote{\url{www.gmrt.ncra.tifr.res.in}} \citep{Paciga2013}, as well as the forthcoming Square Kilometre Array (SKA)\footnote{\url{www.skatelescope.org}} \citep{2015aska.confE...1K}. However, no detection has been made and the derived upper limits set only weak constraints on astrophysical quantities \citep{HERA2021, TheHERACollaboration_2022}.

\subsubsection{The 21 cm power spectrum as a probe of the neutral IGM during the final stages of reionisation}
\begin{figure}
\includegraphics[width=\columnwidth]{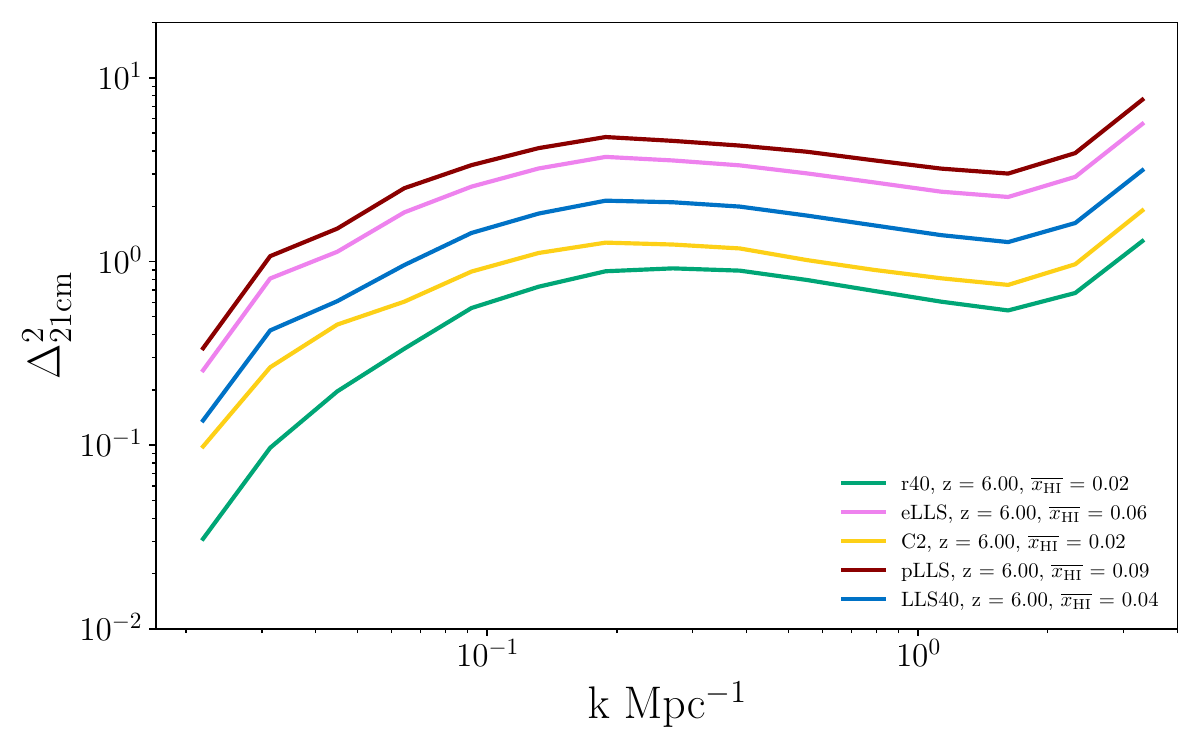}
\includegraphics[width=\columnwidth]{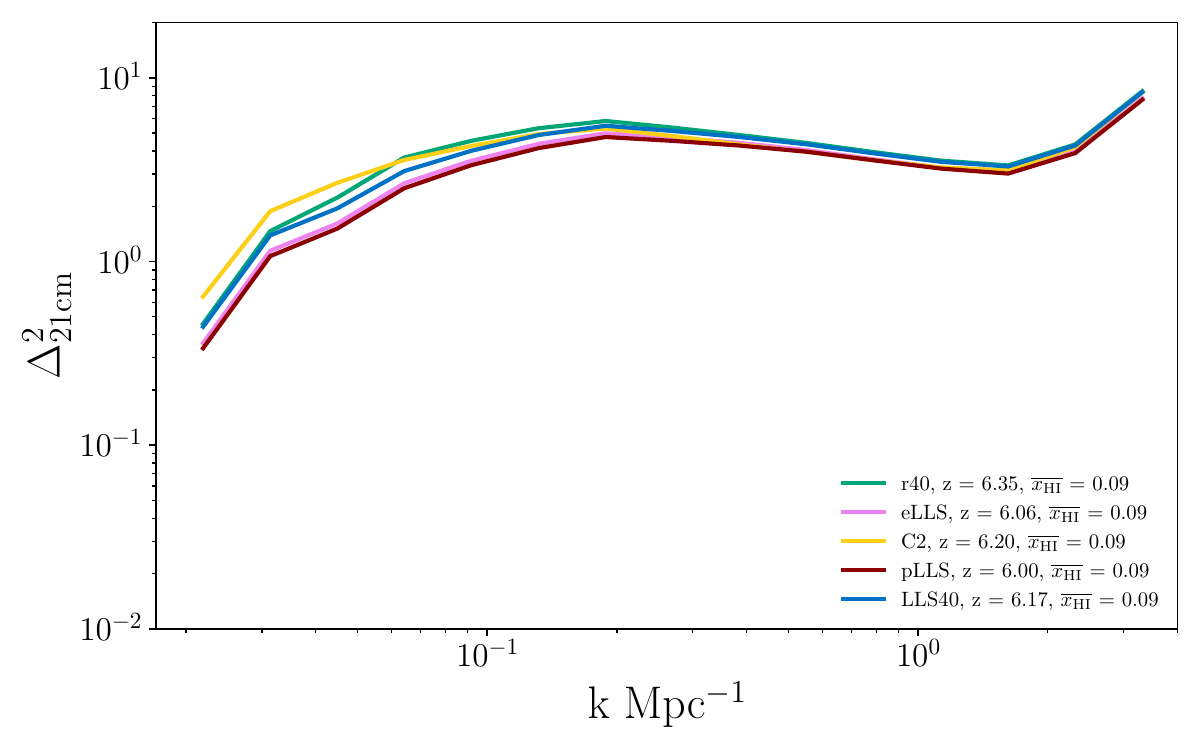}
\caption{The 21 cm power spectrum plots of the models are presented in Tab.~\ref{tab:C2sims} against $k$-scale. The upper panel show the power spectrum at a fixed redshift of 6 for all models. The implementation of the LLS for the diffuse barrier models (\texttt{LLS40, eLLS, pLLS}) results in a delay of the reionisation process. The higher fraction of neutral hydrogen present in each simulation affects the amplitude of the power spectra, accordingly. Note that for the \texttt{r40}, where a cut-off barrier of $R_{\mathrm{max}} = 40$ cMpc is employed across the EoR, results in a steeper slope of the 21 cm power spectrum at low-$k$. The lower panel instead compares each 21 cm power spectrum at a fixed global neutral fraction of 9~\%. The resulting 21 cm power spectra have more or less similar amplitudes due to the fixed global neutral fraction but are sensitive to the distribution of the neutral islands. A large amplitude of the power spectrum at low-$k$ (hence, high physical scale), such as in the \texttt{C2} model, indicates that the majority of the neutral islands in the simulation are fewer and concentrated in the furthest voids (see the lower section of the middle left panel of Fig.~\ref{fig:LC}).}  
\label{fig-ps21}
\end{figure}

In Fig.~\ref{fig-ps21}, we present the 21 cm power spectrum, both at a fixed redshift of 6 (upper panel) and as the closest possible global neutral fraction of around 9~\% (lower panel). The amplitude of the 21 cm power spectrum is notably higher for the diffuse barrier models. Because the diffuse barrier models produce a delayed EoR, more neutral hydrogen is available at redshift 6 (as seen at the bottom $y$-axes of the panels in Fig.~\ref{fig:LC}). This is most apparent when comparing the power spectra at a fixed neutral fraction in the bottom panel of Fig.~\ref{fig-ps21}, where the overall quantity of \HI is similar across all simulations. In doing so, we can reduce the role of the global neutral fraction when comparing our results, and see a quantitative decrease in the ratios. Despite this, the spatial distribution of the neutral islands also plays an important role in the 21 cm power spectrum. 

Taking the \texttt{r40} model, for example, the bubble growth is restricted up to 40 cMpc across all of the EoR. The effect of this is most pronounced when the \HII bubble size reaches the cut-off scale, which in the case for \texttt{r40} is at redshifts $z <6.3$ (seen as the flattening of the UVB amplitude at Sec.~\ref{sec:EoR_Observables}). At a redshift of 6 (the upper panel of Fig.~\ref{fig-ps21}), the low-$k$ 21-cm power spectrum has a steeper slope compared to the other models, as traces the density field, while larger $k$-values are influenced by the ionisation field \citep[see sec.~5 of][]{Georgiev2022}. In comparison, in the \texttt{LLS40} model, where the barrier is also fixed yet diffuse, the transition scale is smoother since 37~\% of the photons (at the Lyman limit) can travel further than the imposed 40 cMpc (see Section~\ref{tab:C2sims}). Note that, in general, both models produce similar results when compared at a similar global neutral fraction (bottom panel of Fig.~\ref{fig-ps21}). Conversely, the \texttt{C2} model does not display a similar feature and indeed has more power on large scales, when compared in the bottom panel of Fig.~\ref{fig-ps21}. 

\begin{figure}
\includegraphics[width=\columnwidth]{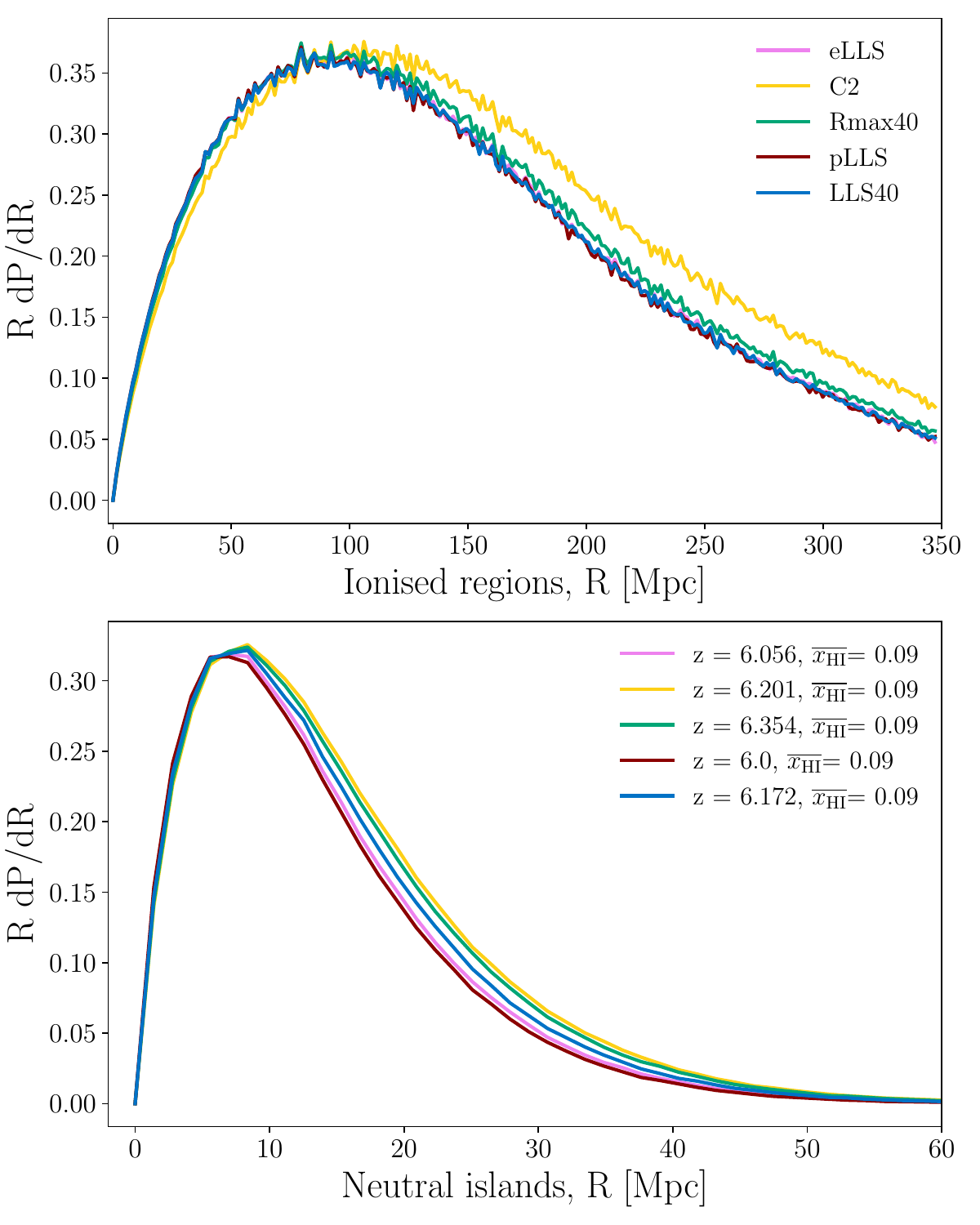}
\caption{Distributions using the Mean-Free-Path size determination method in \citet{Mesinger2007} for at a fixed global neutral fraction. The top panel shows the distribution of ionised \HII regions whilst the bottom panel probes the \HI island sizes. Most notably, both ionised and neutral LOS are slightly longer in the case  of the \texttt{C2} simulation, compared to the other cases from Tab.~\ref{tab:C2sims}.}  
\label{fig-bsd}
\end{figure}

Hence, we have also analysed the same outputs using the MFP-BSD methodology from \citet{Mesinger2007} to probe the \HI island sizes at a fixed global neutral fraction, shown in Fig.~\ref{fig-bsd}. Note that, we confirm that the \texttt{C2} simulation produces slightly larger ionised regions, however, we find the method cannot distinguish between the distribution of the \texttt{r40, LLS40, eLLS, pLLS} models. Therefore, by allowing the photons to freely propagate across the simulations and instead adjusting the recombination factor, the resulting neutral islands at the end of reionisation are mainly larger in size and group at the furthest voids of the simulation, resulting in more power in large scales. 
A similar line of thinking, albeit less pronounced, can be inferred for the \texttt{r40} and \texttt{LLS40} when comparing the sizes of neutral islands in the bottom panel Fig.~\ref{fig-bsd} to the amplitude of the power spectra in the bottom panel of Fig.~\ref{fig-ps21}. The \texttt{eLLS} and \texttt{pLLS} models look largely consistent with one another, as their main difference is centred on how the small-scale absorbers are modelled within the ionised regions. Compared to the remaining models, these diffuse barrier models produce neutral islands which are smaller in size and more scattered across the simulation due to the fluctuations introduced by the LLS. 
\section{Conclusion and discussion}
\label{Sec:Conclusions}

This body of work examines the end stages of the Epoch of Reionisation (EndEoR) using a set of 3D radiative transfer simulations. The final stages of reionisation are particularly difficult to model numerically, due to the significant amount of ionising sources and the complex topology of its ionised overlapping structures. Particularly, modelling the UVB within such simulations often overestimates the measurements reported by \citet{Calverley2011}. Despite this, the EndEoR is a treasure trove of information about the process of reionisation, and the first stars and galaxies. Utilising information from high redshift quasars, the reported low values of the mean free path of ionising photons at a redshift of 6 \citep{Becker2021, Zhu2023} as well as higher values of the effective \lya opacity along the line-of-sight \citep{Bosman2021} indicate at an extended reionisation period \citep{Keating2018,Kulkarni2019}. In addition, the large Gunn–Peterson trough reported by \citet{Becker2015} hints at the presence of islands of neutral hydrogen, which can be utilised to study reionisation through the 21 cm signal \citep{Giri2019}.

This work aims to explore the role of small-scale optically thick self-shielded systems, which are often not included in large-scale simulations of reionisation. Referred to as Lyman Limit systems, these systems are often associated with hydrogen present in the outskirts of galaxies and/or in the filaments between them \citep{Theuns2024,Fan2024}. LLS have an effect on the observables of reionisation, such as the effective \lya opacity, the mean free path of ionising (Lyman continuum) photons, as well as the 21 cm signal from the intergalactic medium. 
We investigate several methods to model these small-scale absorbers with varying assumptions (see Section~\ref{subsec:LLS_models} and Table~\ref{tab:C2sims}). All four {\sc C$^{2}$-Ray} models are identical in set-up save for the modelling of the LLS, described below:
\begin{enumerate}
    \item In the \texttt{r40} model ionising photons are cut off at a fixed distance of 40 cMpc for all redshifts. This can be seen as a model which only considers the large-scale effect of the LLS on reionisation.
    \item The \texttt{LLS40} model uses an additional subgrid opacity to impede the propagation of ionising photons. Following the definition of the mean free path of ionising photons, at 40 cMpc the cumulative additional opacity is equal to unity. Hence, this barrier is uniformly distributed and porous, as roughly 37~\% of the photons (at the Lyman limit) can permeate past the boundary condition. Unlike the \texttt{r40} model, the \texttt{LLS40} will also impact the ionisation rate within the \HII regions. \item The \texttt{eLLS} model uses the same additional opacity approach as \texttt{LLS40} but evolves the value of the mean free path according to the fit from \citetalias{Worseck2014}, see equation~\ref{eq:W14fit}. It thus captures the likely redshift dependence of the number of LLS within the simulation, although calibrated on post-EoR observables.
    \item In the \texttt{pLLS} we use the same MFP fit as in \texttt{eLLS} but distribute the additional opacity according to the a recipe based on the cross-section of dark matter halos, rather than uniformly \citep[see][for a description]{Shukla2016}. This model captures both the likely evolution of the number of LLS as well as the probable connection between LLS and collapsed structures.
    \item Lastly, model \texttt{C2} attempts to model the presence of small-scale absorbers by boosting the global recombination rate by a factor of two. We include this method, as it is an often-used solution within the community.
\end{enumerate}
  
The conclusions of our study can be formulated as follows:

\begin{enumerate}
    \item [1.] We find that by including the presence of LLS in our simulations, we can naturally reproduce delayed reionisation histories. Including redshift and/or density dependence in our LLS models further extends the EoR. Moreover, we quantify how the role of clumping in our \texttt{C2} model results in an overshooting of the UVB as well as how the choice of a cut-off barrier for the \texttt{r40} case can produce inconsistent features in our observables. Conversely, the diffuse barrier models \texttt{LLS40, eLLS, pLLS} are more consistent with the evolution of the UVB reported by high-resolution small-scale simulations of reionisation such as {\sc THESAN} \citep{Garaldi2022} and {\sc CODA III} \citep{Lewis2022}. 
    \item[2.] By conducting a statistical analysis of the UVB for our set of models in Sec.~\ref{sec::UVB_PDF_analysis}, we highlight that the role of Lyman Limit Systems and how they are modelled will affect reionisation globally as well as locally. Models such as \texttt{r40, LLS40}, which only model the impact of the value of the MFP on ionising photons, affect the UVB on the larges-scales (captured by the right side of the PDFs in Fig.~\ref{fig::pdf}) but not the small-scales, where LLS are not directly modelled, and the ionisation rate is higher. Including the redshift evolution of the LLS in \texttt{eLLS} marginally improves on this. Crucially, the density and redshift-dependent \texttt{pLLS} model, which models the LLS on the cell level, results in small-scale fluctuation in the ionised regions, resulting in a distinct log-normal shape of the PDF.
\end{enumerate}

We can summarise that the properties of the Lyman Limit systems affect the global parameters of reionisation. In the last section of the paper, we study whether these findings are consistent with observations at the EndEoR.

\begin{enumerate}
    \item [3.] We conduct mock observations of the effective \lya optical depth ($\tau_{\mathrm{eff}}$) and compare the features of our results against the measurements from the QSO sample from \citetalias{Bosman2022}. In Section.~\ref{subsec:eff_lya_opacity} (see Figure~\ref{fig2}), we confirm our findings in points 1. and 2. necessitating the role of the proper modelling of small-scale absorbers during reionisation. While the \texttt{eLLS,C2} models marginally reproduce the features of the observable, only the \texttt{pLLS} model can match the redshift evolution of the effective \lya optical depth ($\beta = 13.3 \pm 0.6$) from the measurements ($\beta = 13.7 \pm 1.3$).    
    \item [4.] We further our study focusing on direct observations of the mean free path of ionising photons, which as reported by \citet{Becker2021} and \citet{Gaikwad2023} is particularly low at a redshift of 6 and another indicator of a delayed reionisation. In Section \ref{fig:MFP_evolve} we sample lines of sight from our simulations and construct a PDF of the free paths of ionising photons at a redshift of 6 or a fixed neutral fraction of 15~\% (see Fig.~\ref{fig:MFP}). We find that models such \texttt{r40} and \texttt{C2} overshoot the value of the mean free path as their implementations and cannot induce the required opacity to limit Lyman limit photons. And while all diffuse barrier models (\texttt{LLS40, eLLS, pLLS}) reproduce low values of the MFP at a redshift of 6 $\lambda_{\mathrm{MFP}}\approx 25$ cMpc, their distributions differ massively due to the nature of their implementations. Notably, the \texttt{pLLS} model follows a log-normal distribution skewed towards low values of the MFP, and both redshift-dependent absorption models (\texttt{pLLS} and \texttt{eLLS}) have distributions with standard deviations overlapping within the MFP observations (see Fig.~\ref{fig:MFP_evolve}).
    \item[5.] Lastly, we probe the effect of Lyman Limit Systems on the neutral medium during EndEoR by examining the 21 cm signal, particularly the 21 cm power spectrum as shown in Fig.~\ref{fig-ps21}. At redshift 6, we find differences in amplitude by an order of magnitude between models such \texttt{r40} and \texttt{pLLS}, which can be attributed to the delay of the reionisation process discussed in 1. Note, the flat shape of the \texttt{r40} is consistent with the discussion in \citet{Georgiev2022}. When comparing the 21 cm power spectra at a fixed neutral fraction of 9~\%, we can further peer into key differences between models. We find that the clumping model has a higher large-scale amplitude of the 21 cm signal, indicative of larger neutral islands. Since the clumping model has no imposed MFP, the majority of neutral islands trace the furthest voids from the clusters of ionised sources. And while the \texttt{LLS40} model imposes a barrier of 40 cMpc, its permissivity allows for ionising photons to traverse past it, resulting in a similar effect on the neutral islands. Conversely, the \texttt{r40} is seen to have power across all $k$-scales, because most neutral islands will be located in regions 40 cMpc away from ionising sources.
    
    Lastly, both \texttt{eLLS} and \texttt{pLLS} have a similar shape and amplitude. This occurs because the \HI islands are less sensitive to the modelling differences of the small-scale absorbers within the ionised bubbles, but rather on the large-scale mean free path imposed by the LLS and its redshift evolution.
\end{enumerate}

In summary, we find that modelling the LLS during the epoch of reionisation has a crucial role in regulating the duration of reionisation as well its observables such as the UVB background, \lya effective optical depth, mean free path of ionising photons as well as the shape of neutral islands detentions via the 21 cm signal. In concordance with \citet{Georgiev2022} we conclude that cut-off barrier models such as \texttt{r40}, which only place an upper limit on the MFP value, produce inconsistent results with the current knowledge within the field. Meanwhile, the additional clumping added in the \texttt{C2} allows the model to marginally reproduce some of the features from the $\tau_{\mathrm{eff}}$ observations. However, it is insufficient in reducing the average distance of UV photons, resulting in higher values of the UVB and MFP at redshift 6, compared to observations. This also impacts the 21-cm power spectrum, which when compared to all models at a fixed neutral fraction, has more power on low-$k$ scales. Introducing a uniform opacity diffuse barrier like in \texttt{LLS40} resolves many of the aforementioned problems, but is not fully consistent with observations of the evolution of the \lya optical depth observations. Lastly, the
evolving diffuse barrier model \texttt{eLLS} improves on the latter and reproduces the observables adequately, but it is only when we account for the density-dependence of the absorbers such as in \texttt{pLLS} is sufficient to reproduce the measurements of $\tau_{\mathrm{eff}}$ and the MFP. When comparing the 21-cm power spectrum at $\xb \approx 0.09$, both \texttt{eLLS} and \texttt{pLLS} are identical, while \texttt{LLS40} is shown to match the \texttt{r40} model and have more power on low-$k$ scales. The addition of a redshift and position dependence to the model of small-scale absorbers is shown to delay the EndEoR (comparing \texttt{LLS40,eLLS,pLLS} respectively).

Although we covered a wide range of approaches for including the effect of unresolved absorbers, there remain further models, to investigate. For example, the evolving hard barrier model from \citet{giri2024end} can be combined with the halo model in \citet{Theuns2024}, which reproduces the MFP evolution from \citetalias{Worseck2014}, although the disadvantages of the hard barrier approach as shown in the current paper will
also be present in this model. A more promising approach is the density-dependent clumping models reported in \citet{Sobacchi2014} and \citet{Bianco2021} as well as that of \citet{Cain2021}, where the small-scale absorbers are included via sub-grid model informed by small-scale high-resolution simulations. Such approaches, through the application of clumping factors, can capture the dynamic response of IGM clumping to the UVB in a fully self-consistent manner and are a promising avenue for future work. We will investigate the sensitivity of the parameters of this fit for models such as  \texttt{pLLS}. Moreover, we will extend our simulations to a redshift of $z \approx 5.5$ to further explore the EndEoR, however, and explore the role of our source model on the duration of reionisation \citep[e.g.][]{Cain2023b}.

\section*{Acknowledgements}
We thank the anonymous referee for constructive and insightful comments. IG acknowledges the support from the Gustaf och Ellen Kobbs stipendiestiftelse and the Alva and Lennart Dahlmark Research Grant. GM’s research is supported by the Swedish Research Council project grant 2020-04691\_VR. Nordita is supported in part by NordForsk. Other results were obtained on resources provided by the National Academic Infrastructure for Supercomputing in Sweden (NAISS) at PDC (DARDEL), Royal Institute of Technology, Stockholm. We have utilised the following python packages for manipulating the simulation outputs and plotting results: numpy \citep{harris2020array}, scipy \citep{2020SciPy-NMeth}, matplotlib \citep{Hunter2007} and{ \sc tools21cm}\footnote{\url{https://github.com/sambit-giri/tools21cm}} package \citep{Giri2020}.
\section*{Data Availability}
The data underlying this article will be shared on a reasonable request
to the corresponding author.



\bibliographystyle{mnras}
\bibliography{example} 




\appendix
\bsp	
\label{lastpage}
\end{document}